\documentclass[graybox]{svmult}

\usepackage{mathptmx}       
\usepackage{helvet}         
\usepackage{courier}        
\usepackage{type1cm}        
\usepackage{makeidx}         
\usepackage{graphicx}        
\usepackage{multicol}        
\usepackage[bottom]{footmisc}

\usepackage{color}

\usepackage{amsmath,amssymb}
\usepackage{braket}
\usepackage{cite}

\makeindex             

\begin{document}

\title*{Time-Dependent Complete-Active-Space Self-Consistent-Field Method for Ultrafast Intense Laser Science}
\titlerunning{TD-CASSCF Method for Ultrafast Intense Laser Science}
\author{Takeshi Sato, Yuki Orimo, Takuma Teramura, Oyunbileg Tugs, and Kenichi L. Ishikawa}
\institute{Takeshi Sato \at Department of Nuclear Engineering and Management, Graduate School of Engineering, The University of Tokyo, 7-3-1 Hongo, Bunkyo-ku, Tokyo 113-8656, Japan, \email{sato@atto.t.u-tokyo.ac.jp}
\and Yuki Orimo \at Department of Nuclear Engineering and Management, Graduate School of Engineering, The University of Tokyo, 7-3-1 Hongo, Bunkyo-ku, Tokyo 113-8656, Japan, \email{ykormhk@atto.t.u-tokyo.ac.jp}
\and Takuma Teramura \at Department of Nuclear Engineering and Management, Graduate School of Engineering, The University of Tokyo, 7-3-1 Hongo, Bunkyo-ku, Tokyo 113-8656, Japan, \email{teramura@atto.t.u-tokyo.ac.jp}
\and Oyunbileg Tugs \at Department of Nuclear Engineering and Management, Graduate School of Engineering, The University of Tokyo, 7-3-1 Hongo, Bunkyo-ku, Tokyo 113-8656, Japan, \email{tugs@atto.t.u-tokyo.ac.jp}
\and Kenichi L. Ishikawa \at Department of Nuclear Engineering and Management, Graduate School of Engineering, The University of Tokyo, 7-3-1 Hongo, Bunkyo-ku, Tokyo 113-8656, Japan, \email{ishiken@n.t.u-tokyo.ac.jp}
}
%
\maketitle


\abstract{We present the time-dependent complete-active-space self-consistent-field (TD-CASSCF) method to simulate multielectron dynamics in ultrafast intense laser fields from the first principles.
While based on multiconfiguration expansion, it divides the orbital space into frozen-core (tightly bound electrons with no response to the field), dynamical-core (electrons tightly bound but responding to the field), and active (fully correlated to describe highly excited and ejected electrons) orbital subspaces.
The subspace decomposition can be done flexibly, conforming to phenomena under investigation and desired accuracy.
The method is gauge invariant and size extensive.
Infinite-range exterior complex scaling in addition to mask-function boundary is adopted as an efficient absorbing boundary.
We show numerical examples and illustrate how to extract relevant physical quantities such as ionization yield, high-harmonic spectrum, and photoelectron spectrum from our full-dimensional implementation for atoms.
The TD-CASSCF method will open a way to the {\it ab initio} simulation study of ultrafast intense laser science in realistic atoms and molecules. 
}

\section{Introduction}
\label{sec:Introduction}

From atoms and molecules under visible-to-midinfrared laser fields of an intensity $\gtrsim 10^{14}\,{\rm W/cm}^2$ emerge highly nonlinear strong-field phenomena, e.g., above-threshold ionization, tunneling ionization, high-harmonic generation (HHG), and nonsequential double ionization (NSDI) \cite{Protopapas1997RPP,Brabec2000RMP}. 
In particular, HHG is more and more widely used as an ultrashort (down to attoseconds) coherent light source in the extreme-ultraviolet (XUV) and soft x-ray spectral ranges \cite{Popmintchev2012Nature,Chang2011,AttosecondPhysics}. 
In addition, free-electron lasers are now in operation as another type of ultrashort, intense, coherent XUV and x-ray sources.
Such a rapid progress in experimental techniques for ultrafast intense laser science has opened new research areas including ultrafast molecular probing
\cite{Itatani2004Nature,Haessler2010NPhys,Salieres2012RPP}, attosecond science \cite{Agostini2004RPP,Krausz2009RMP,Gallmann2013ARPC}, and XUV nonlinear optics \cite{Sekikawa2004Nature,Nabekawa2005PRL}, with the ultimate goal to directly observe, and even manipulate ultrafast electronic motion in atoms, molecules, and solids.

Further advances in these areas require first-principles methods to numerically simulate the real-time dynamics of multielectron atoms and molecules in ultrafast intense laser pulses, or {\it ab initio strong-field physics}. 
Although the time-dependent Schr\"odinger equation (TDSE) [see Eq.~(\ref{eq:TDSE}) below] rigorously describes these phenomena in principle,
its numerical integration in the real space for systems with more than two electrons \cite{Pindzola1998PRA,Pindzola1998JPB,Colgan2001JPB,
Parker2001,Laulan2003PRA,Piraux2003EPJD,Laulan2004PRA,ATDI2005,
Feist2009PRL,Pazourek2011PRA,He_TPI2012PRL,Suren2012PRA,He_TPI2013AS,
Vanroose2006PRA,Horner2008PRL,Lee2010JPB} poses a major challenge.

A promising class of approaches is time-dependent multiconfiguration self-consistent field (TD-MCSCF) methods\index{Time-dependent multiconfiguration self-consistent field (TD-MCSCF) method}\cite{Ishikawa2015JSTQE,Loestedt2017PUILS}, where the total electronic wave function is expressed as a superposition of different electronic configurations or Slater determinants built from a given number of single-electron spin orbitals [see Eq.~(\ref{eq:general-mcwf}) and Fig.~\ref{fig:MC} below].
In the multiconfiguration time-dependent Hartree-Fock (MCTDHF) method\index{Multiconfiguration time-dependent Hartree-Fock (MCTDHF) method}\cite{Zanghellini2003LP,Kato2001CPL,Caillat2005PRA}, both the expansion coefficients [configuration-interaction (CI) coefficients] and orbital functions are varied in time, and all the possible realizations to distribute the electrons among the spin orbitals (full CI expansion) are included.
Though pioneering and powerful, the computational cost of MCTDHF increases factorially with the number of electrons.

To overcome this difficulty, we have recently developed and successfully implemented a TD-MCSCF method called the time-dependent complete-active-space self-consistent-field (TD-CASSCF) method \cite{Sato2013PRA,Sato2016PRA,Orimo2018PRA}, which is the topic of the present Chapter. 
TD-CASSCF classifies the spatial orbitals into doubly occupied and time-independent frozen core (FC), doubly occupied and time-dependent dynamical core (DC), and fully correlated active orbitals.
Thanks to this classification, the number of configurations used in simulations and the computational cost are significantly reduced without sacrificing accuracy.
The classification can be done flexibly, based on simulated physical situations and desired accuracy. 
Through comparison of the results from various subspace decompositions, one can analyze the contribution from different shells, the effect of electron correlation, and the mechanism underlying the simulated phenomena.
In this sense, TD-CASSCF is even more useful than merely numerically exact black-box simulations.

This Chapter proceeds as follows. In Sec.~\ref{sec:Problem Statement} we describe the statement of the problem that we are going to treat, i.e., the time-dependent Schr\"odinger equation for many electron systems in a driving laser field within the dipole and fixed-nuclei approximations.
We also briefly mention an important concept of gauge transformation.
Section \ref{sec:TD-CASSCF} explains the formulation of the TD-CASSCF method, the equations of motion for CI coefficients and orbital functions, and its important features of gauge invariance and size extensivity.
In Sec.~\ref{sec:Initial-State Preparation and Simulation Boundary} we describe how to prepare the initial wave function and absorb the electron wave packet that reaches the simulation box boundary without unphysical reflection.
Section \ref{sec:Numerical examples} presents how to extract relevant physical quantities from the wave function obtained by TD-CASSCF simulations, along with representative numerical examples.
Summary is given in Sec.~\ref{sec:Summary}.
Hartree atomic units are used throughout unless otherwise stated.

%

\section{Problem Statement}
\label{sec:Problem Statement}

\subsection{Time-Dependent Schr\"odinger Equation}

We consider an atom or molecular system consisting of $N$ electrons subject to an external laser field. Within the electric dipole approximation of laser-electron interaction and the fixed-nuclei or clamped-nuclei approximation that treats nuclei as classical point charges fixed in space, the dynamics of the laser-driven multielectron system is described by the time-dependent Schr\"odinger equation (TDSE)\index{Time-dependent Schr\"odinger equation (TDSE)},
\begin{equation}
\label{eq:TDSE}
i\frac{\partial\Psi (t)}{\partial t} = \hat{H}(t)\Psi (t),
\end{equation}
where the time-dependent Hamiltonian,
\begin{equation}
\hat{H}(t)=\hat{H}_1(t)+\hat{H}_2,
\end{equation}
is decomposed into the one-electron part (kinetic energy, nuclear Coulomb energy, and laser-electron interaction),
\begin{equation}
\label{eq:H1}
\hat{H}_1(t) = \sum_i \hat{h}({\bf r}_i,t) 
\end{equation}
and the two-electron part,
\begin{equation}
\hat{H}_2 = \sum_{i=1}^N \sum_{j < i} \frac{1}{|{\bf r}_i - {\bf r}_j|},
\end{equation}
for the interelectronic Coulomb interaction. 
The laser-electron interaction can be expressed either in the length gauge (LG)\index{Length gauge} or velocity gauge (VG)\index{Velocity gauge}: $\hat{h}({\bf r},t)$ in Eq.~(\ref{eq:H1}) is given by,
\begin{equation}
\label{eq:length-gauge}
\hat{h}({\bf r},t) = \frac{\hat{{\bf p}}^2}{2}+{\bf r}\cdot {\bf E}(t)-\sum_\alpha \frac{Z_\alpha}{|{\bf r} - {\bf R}_\alpha|},
\end{equation}
in the length gauge, with $\hat{{\bf p}}=-i\nabla$, and, 
\begin{equation}
\label{eq:velocity-gauge}
\hat{h}({\bf r},t) = \frac{\left[\hat{{\bf p}}+{\bf A}(t)\right]^2}{2}-\sum_\alpha \frac{Z_\alpha}{|{\bf r} - {\bf R}_\alpha|},
\end{equation}
in the velocity gauge, with ${\bf A}(t) = -\int {\bf E}(t)dt$ being the vector potential.

\subsection{Gauge Transformation}
\label{subsec:Gauge Transformation}

The wave functions $\Psi_{\rm L} (t)$ and $\Psi_{\rm V} (t)$ expressed in the length and velocity gauges, respectively, are transformed into each other through the gauge transformation\index{Gauge transformation},
\begin{equation}
\label{eq:gauge-transformation}
\Psi_{\rm V}(t) = \hat{\mathcal{U}}(t) \Psi_{\rm L}(t),
\end{equation}
with the unitary operator,
\begin{equation}
\label{eq:gauge-transformation-operator}
\hat{\mathcal{U}}(t) = \exp \left[-i{\bf A}(t)\cdot \sum_{i=1}^{N}{\bf r}_i\right].
\end{equation}
If we substitute Eq.~(\ref{eq:gauge-transformation}) into the TDSE with Eq.~(\ref{eq:velocity-gauge}), we can easily show that $\Psi_{\rm L}$ indeed satisfies the TDSE with Eq.~(\ref{eq:length-gauge}).

While the operator $\hat{{\bf p}}$ corresponds to the kinetic momentum in the length gauge, it corresponds to the canonical momentum in the velocity gauge, and the kinetic momentum is given by $\hat{{\bf p}}+{\bf A}(t)$. Then, a plane wave state with a kinetic momentum ${\bf p}_{kin}$ is $e^{i {\bf p}_{kin} \cdot {\bf r}}$ in the length gauge and $e^{i \left[{\bf p}_{kin}-{\bf A}(t)\right] \cdot {\bf r}}$ in the velocity gauge, which fulfills Eq.~(\ref{eq:gauge-transformation}).

The gauge principle\index{Gauge principle} states that all physical observables are gauge invariant\index{Gauge invariance}, i.e., take the same values whether the laser-electron interaction may be represented in the length or velocity gauge \cite{Bandrauk2013JPB}.
For example, the probability density is gauge invariant, $|\Psi_{\rm V}(t)|^2 = |\Psi_{\rm L}(t)|^2$.

One may be surprised to realize that the projection $\langle \Xi | \Psi (t) \rangle$ of the wave function $\Psi (t)$ onto a field-free stationary state $\Xi$ and the population $|\langle \Xi | \Psi (t) \rangle|^2$ are {\it not} gauge invariant and, thus, {\it not} a physical observable when ${\bf A}(t) \ne 0$, i.e., during the pulse. 
As a consequence, the degree of ionization is {\it not} gauge invariant during the pulse, either.
Let us assume that a hydrogen atom under a laser field linearly polarized in the $z$ direction is in the ground state in the length gauge,
\begin{equation}
	\psi_{\rm L}({\bf r},t) = \frac{e^{-r}}{\sqrt{\pi}},
\end{equation}
at some moment, e.g., after a complete Rabi oscillation cycle.
Then, its velocity gauge wave function is,
\begin{equation}
\label{eq:HgsVG}
	\psi_{\rm V}({\bf r},t) = e^{-iA(t)z} \psi_{\rm L}({\bf r},t)
	= 2 e^{-r}\sum_{l=0}^\infty \sqrt{2l+1}(-i)^lj_l(A(t)r)Y_{l0}(\theta,\phi),
\end{equation}
which contains not only the $1s$ state but all the angular momenta $l$ including continuum levels unless $A(t)=0$.

\section{TD-CASSCF method}
\label{sec:TD-CASSCF}

\index{Time-dependent complete-active-space self-consistent-field (TD-CASSCF) method}

\subsection{Multicongifuration Expansion}
\label{subsec:Multicongifuration Expansion}

\begin{figure}[tb]
\begin{center}
\includegraphics[width=0.75\textwidth]{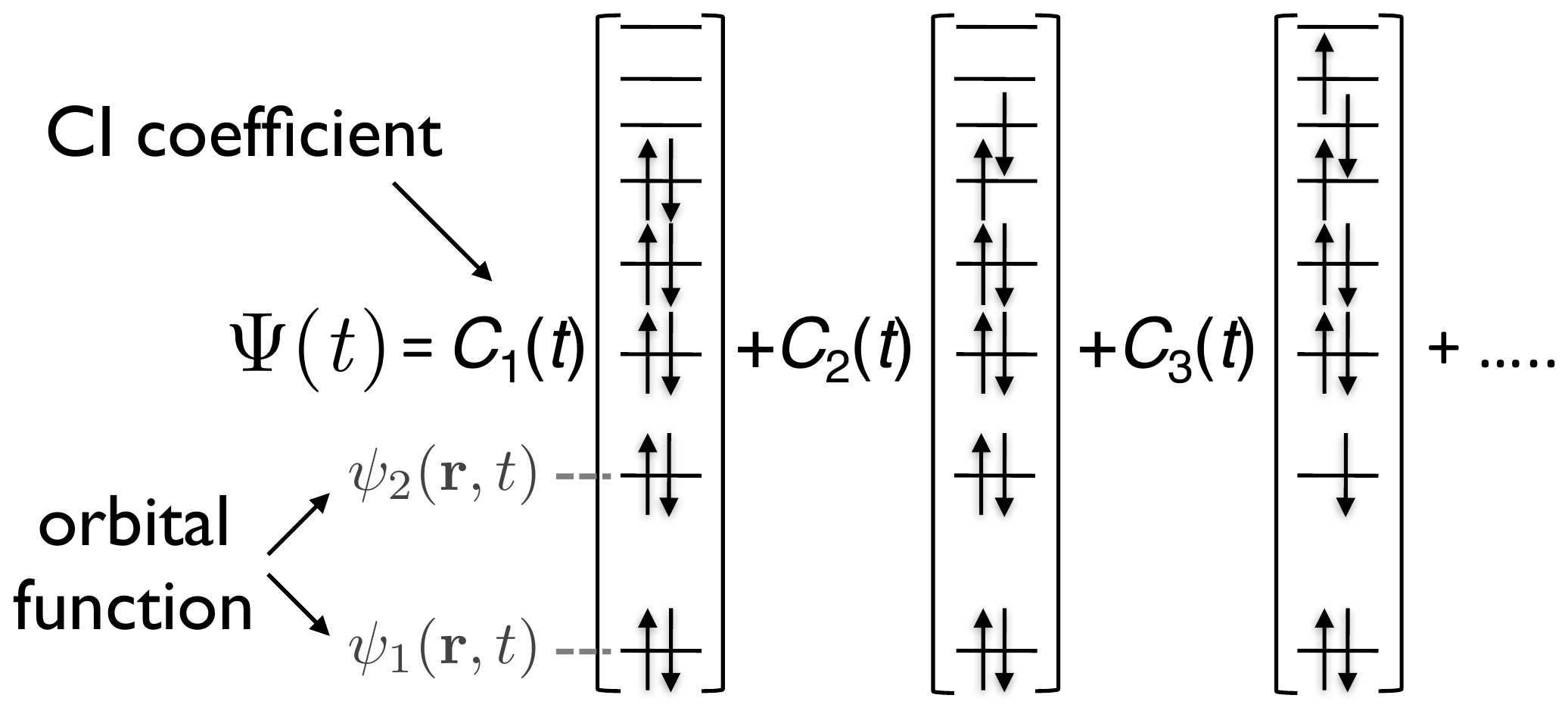}	
\end{center}
%
%
\caption{Schematic representation of the multiconfiguration expansion Eq.~(\ref{eq:general-mcwf}). Each term on the right-hand side corresponds to a configuration $\Phi_1, \Phi_2, \cdots$ with CI coefficients $C_1, C_2, \cdots$. The first term corresponds to the Hartree-Fock configuration.}
\label{fig:MC}       
\end{figure}

In order to simulate multielectron dynamics, as illustrated in Fig.~\ref{fig:MC}, we expand the total wave function $\Psi (t)$ as a superposition of different Slater determinants or configuration state functions\index{Multiconfiguration expansion},
\begin{equation}
\label{eq:general-mcwf}
\Psi (t) = \sum_{I}^{\sf P} \Phi_{I}(t)C_{I}(t),
\end{equation}
where expansion coefficients $\{C_{I}\}$ are called configuration interaction (CI) coefficients\index{Configuration interaction coefficient}\index{CI coefficient} and bases $\{\Phi_{I}\}$ are the Slater determinants  built from $N$ spin orbitals out of $2n$ spin orbitals $\{\psi_p (t);p=1,2,\cdots,n\}\otimes\{\alpha,\beta\}$  (in the spin-restricted treatment) with $\{\psi_p\}$ being spatial orbital functions and $\alpha (\beta)$ the up- (down-) spin eigenfunction. The summation in Eq.~(\ref{eq:general-mcwf}) with respect to configurations $I$ runs through the
element of a CI space ${\sf P}$, consisting of
a given set of determinants. 

Muticonfiguration expansion Eq.~(\ref{eq:general-mcwf}) can represent a wide variety of different methods; whereas $\{C_I\}$ are usually taken as time-dependent, they can also be fixed \cite{Miranda2011JCP}. $\{\psi_p\}$, and thus $\{\Phi_I\}$, can be considered either time-independent, as in the time-dependent configuration interaction singles (TDCIS) method \cite{Greenman_2010}, or time-dependent, as in the TD-CASSCF, MCTDHF, and time-dependent Hartree-Fock (TDHF) \cite{Pindzola1991PRL} methods described below. 
While orbital functions are usually assumed to fulfill orthonormality, it is not a necessary condition.

\subsection{TD-CASSCF ansatz}

In the TD-CASSCF method, we use orthonormal time-dependent orbital functions. The $n$ occupied orbitals are classified into $n_c$ core orbitals\index{Core orbitals} $\{\psi_i: i=1,2,\cdot\cdot\cdot,n_c\}$ that are doubly occupied all the time and $n_a(=n-n_c)$ active orbitals\index{Active orbitals} $\{\psi_t: t=n_c+1,n_c+2,\cdot\cdot\cdot,n\}$.
This idea is based on a reasonable expectation that only high-lying electrons are strongly driven, while deeply bound core electrons remain nonionized. 
On the other hand, we consider all the possible distributions of $N_a(=N-2n_c)$ electrons among $n_a$ active orbitals. It should be noticed that not only the active orbitals but also the core orbitals, though constrained to the closed-shell structure, vary in time, in general, responding to the field formed by the laser and the other electrons.
The use of time-dependent (especially active) orbitals that are initially localized near the nuclei but spatially expand in the course of time allows us to efficiently describe excitation and ionization.

\begin{figure}[tb]
\begin{center}
\includegraphics[width=\textwidth]{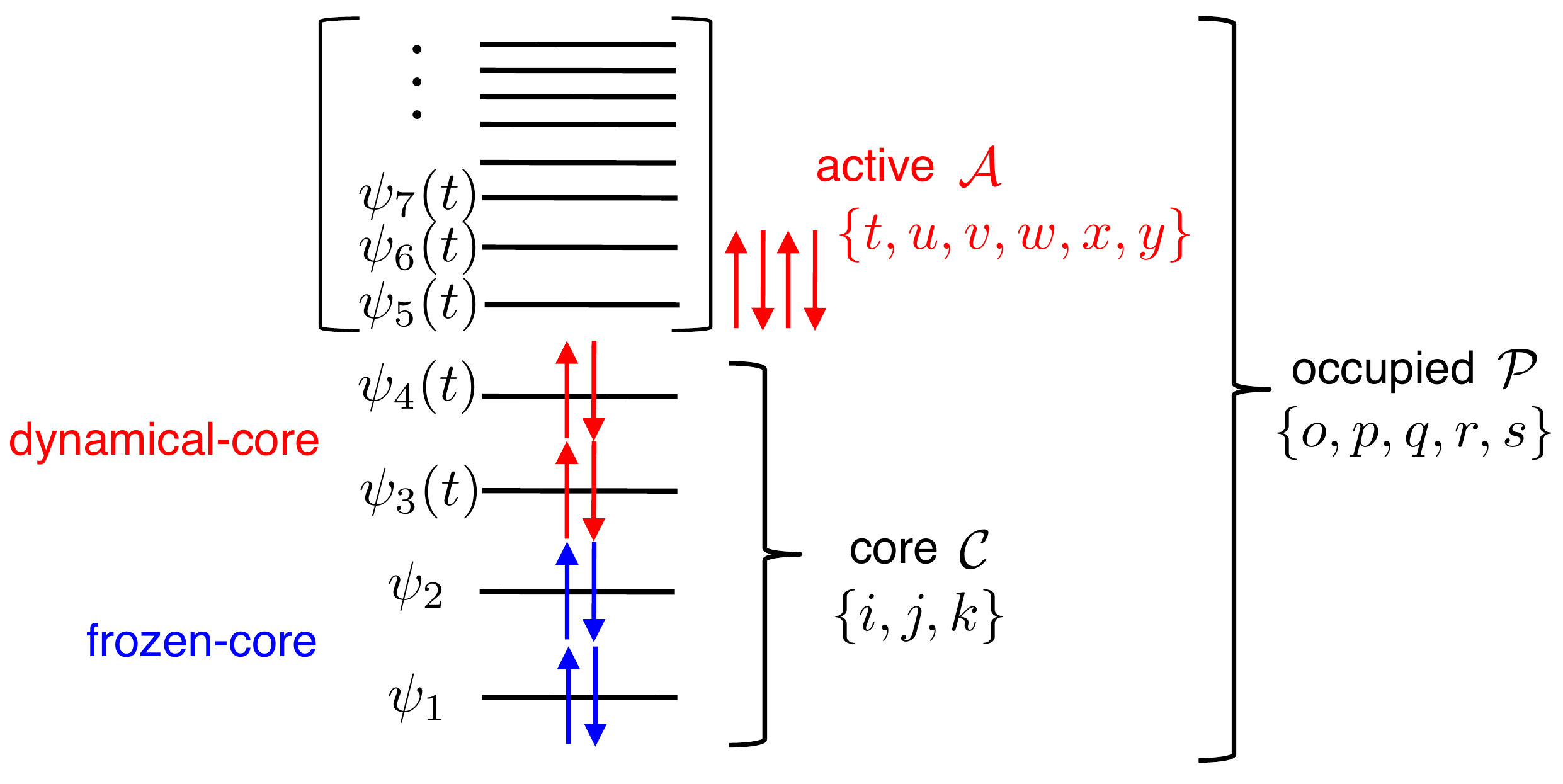}	
\end{center}
%
%
\caption{Schematic illustration of the TD-CASSCF concept for a twelve-electron system with two frozen-core, two dynamical-core, and eight active orbitals. The classification of orbitals and the indices we use are also shown.}
\label{fig:CASSCF-concept}       
\end{figure}

It is also possible to further decompose core orbitals into $n_{fc}$ frozen-core (FC) orbitals\index{Frozen-core (FC) orbitals} that do not vary in time and $n_{dc}$ time-dependent dynamical core (DC) orbitals ($n_c = n_{fc} + n_{dc}$)\index{Dynamical-core (DC) orbitals}. The $N$-electron CASSCF wave function can be symbolically expressed as,
\begin{align}
\Psi_{\rm CAS} : \psi_1^2\cdots\psi_{n_{fc}}^2\psi_{n_{fc}+1}^2(t)\cdots\psi_{n_{c}}^2(t) \{\phi_{n_c+1}(t)\cdots\phi_{n}(t)\}^{N_A},
\end{align}
and given by,
\begin{eqnarray}\label{eq:casscf}
\Psi_\textrm{CAS} = \hat{A}\left[\Phi_\textrm{fc}\Phi_\textrm{dc}(t)\sum_I \Phi_I(t) C_I(t)\right],
\end{eqnarray}
where $\hat{A}$ is the antisymmetrization operator, $\Phi_\textrm{fc}$ and $\Phi_\textrm{dc}$ are the closed-shell
determinants constructed with FC and DC orbitals, respectively, and
$\{\Phi_I\}$ are the determinants formed {\color{black}by} active orbitals.
In the following, we will denote the level of the CAS approximation
employed in $\Psi_{\rm CAS}$
by the integer triple  $(n_{fc}, n_{dc}, n_{a})$.
Hereafter, we use orbital indices $\{i,j,k\}$ for core ($\mathcal{C}$),
$\{t,u,v,w,x,y\}$ for active ($\mathcal{A}$), and $\{o,p,q,r,s\}$ for arbitrary occupied (core and active)
($\mathcal{P}=\mathcal{C}+\mathcal{A}$)
orbitals (Fig.\ \ref{fig:CASSCF-concept}).
The FC and DC orbitals are distinguished explicitly only when necessary.

There are two limiting cases. On one hand, if we use a single configuration made up of only DC orbitals, i.e., $(0,N/2,0)$, or equivalently $(0,0,N/2)$, it corresponds to TDHF\index{Time-dependent Hartree-Fock (TDHF) method}\cite{Pindzola1991PRL}, where some orbitals can also be frozen in a broader sense.
On the other hand, the special case $(0,0,n)$ ($n>N/2$), where all the orbitals are fully correlated or treated as active, corresponds to MCTDHF.

\subsection{TD-CASSCF equations of motion}

The equations of motion (EOMs) that govern the temporal evolutions of the CI coefficients $\{C_I(t)\}$ and orbital functions $\{\psi_p(t)\}$ have been derived on the basis of the time-dependent variational principle (TDVP)\index{Time-dependent variational principle (TDVP)}\cite{Frenkel,Loewdin1972CPL,Moccia1973IJQC}, which requires the action integral,
\begin{equation}
\label{eq:action-integral}
S[\Psi] = \int_{t_0}^{t_1}\langle\Psi | \left(\hat{H}-i\frac{\partial}{\partial t}\right)|\Psi\rangle,
\end{equation}
to be stationary, i.e.,
\begin{equation}
\label{eq:TDVP}
\delta S = \delta\langle\Psi|\hat{H}|\Psi\rangle-i\left(\langle\delta\Psi | \frac{\partial\Psi}{\partial t}\rangle
-\langle\frac{\partial\Psi}{\partial t}|\delta\Psi \rangle\right) = 0,
\end{equation}
with respect to arbitrary variation of CI coefficients and orbitals. By substituting Eq.~(\ref{eq:casscf}) into Eq.~(\ref{eq:TDVP}) and after laborious algebra, one can derive the equations of motion for the CI coefficients and orbital functions.

The form of the resulting EOMs {\color{black}is} not unique
but can be written in various equivalent ways \cite{Sato2016PRA}. Here we present 
the EOMs in the form convenient for numerical implementation \cite{Sato2016PRA}. 
The EOMs for the CI coefficients read,
\begin{eqnarray}
\label{eq:eom_split_cic}
i \frac{d}{dt}C_I (t) &=& \sum_J \langle \Phi_I|\hat{H}_2|\Phi_J\rangle C_J (t),
\end{eqnarray}
which describes transitions among different configurations solely mediated by the interelectronic Coulomb interaction.
The EOMs of the orbitals are given by
\begin{eqnarray}
\label{eq:eom_split_orb}
i\frac{d}{dt}|{\psi}_p\rangle &=& \hat{h}|\psi_p\rangle +\hat{Q} 
\hat{F} |\psi_p\rangle + \sum_q |\psi_q\rangle R^q_p,
\end{eqnarray}
where $\hat{Q}=1-\sum_p |\psi_p\rangle\langle\psi_p|$ is the projector
onto the orthogonal complement of the occupied orbital space. $\hat{F}$ is a mean-field operator that describes the contribution from the interelectronic Coulomb interaction, defined by
\begin{align}\label{eq:fock2}
\hat{F} |\psi_p\rangle = \sum_{oqsr} (D^{-1})_p^o P^{qs}_{or} \hat{W}^r_s |\psi_q\rangle,
\end{align}
where {\color{black} $D$} and {\color{black} $P$} are the one- and two-electron reduced
density matrix (RDM) in the orbital representation, respectively
(see Ref.~\cite{Sato2013PRA} for their definition and the simplification
due to the core-active separation), 
and $\hat{W}^r_s$ is the electrostatic potential of an orbital product (pair potential),
\begin{eqnarray}\label{eq:meanfield}
\hat{W}^r_s({\bf r}) &=&
\int d{{\bf r}^\prime}
\frac{\psi^*_r({{\bf r}^\prime}) \psi_s({{\bf r}^\prime})}
{|{\bf r} - {\bf r}^\prime|}.
\end{eqnarray}
The matrix element $R^q_p$,
\begin{eqnarray}\label{eq:nonredundent}
R^q_p \equiv i \langle\psi_q|\dot{\psi}_p\rangle-h^q_p,
\end{eqnarray}
with $h^q_p = \braket{\psi_q|\hat{h}|\psi_p}$, determines the components of the time derivative of orbitals
{\color{black}in the subspace spanned by the occupied orbitals}.
The elements within one subspace, i.e.{\color{black},} $R^i_j$ and $R^u_t$, can be
arbitrary Hermitian matrix elements and are set to zero $R^i_j=R^u_t=0$
in our implementation \cite{Sato2016PRA}. The elements between the
core and active subspaces are given by,
\begin{align}
	\label{eq:R_ti}
		&R^{t}_i = \left(R^i_t\right)^* =  \left\{
\begin{array}{cc}
{-h^t_i} & {\rm (LG)}\\
{-h^t_i} - \vec{E}(t) \cdot \vec{r}^{\,t}_{\,i} &  {\rm (VG)}\\
\end{array}
\right.& (\text{for } i \in \text{frozen core}), \\
		\label{eq:R1}
		&R^{t}_i = \left(R^i_t\right)^*  =  \sum_u [(2-D)^{-1}]^{t}_{u} (2F^u_i - \sum_v D^u_v F^{i*}_v) 
		& (\text{for } i \in \text{dynamical core}), 
\end{align}
where $F^u_i = \langle\psi_u | \hat{F} | \psi_i \rangle$, and $\vec{r}^{\,t}_{\,i}$ denotes a matrix element of the position vector $\vec{r}$.
For the sake of gauge invariance (see Sebsec.~\ref{subsec:Gauge Invariance}), frozen core orbitals, which are time-independent in the length gauge, are to be varied in time in the velocity gauge as \cite{Sato2016PRA},
\begin{equation}
\psi_i( \vec{r}, t)  =  
	 {\rm e}^{- i \vec{A}(t) \cdot \vec{r}} \psi_i( \vec{r}, 0) \qquad (\text{for } i \in \text{frozen core}),
\end{equation}
in spite of their name. Nevertheless, the FC orbital electron density distribution $|\psi_i( \vec{r}, t)|^2=|\psi_i( \vec{r}, 0)|^2$ is still time-independent.

It is noteworthy that the laser-electron interaction is explicitly contained only in the first term of the orbital EOM Eq.~(\ref{eq:eom_split_orb}) and does not directly drive temporal change of the CI coefficients in Eq.~(\ref{eq:eom_split_cic}).
Thus, in the form presented here, we can say that dynamical correlation induced by the laser field manifests itself first in the orbital EOMs and then spreads to the CI coefficients via the temporal change of orbitals (and, thus, of Slater determinants) in Eq.~(\ref{eq:eom_split_cic}).

\subsection{Numerical Implementation for Atoms}
\label{subsec:Numerical Implementation for Atoms}

We have recently numerically implemented the TD-CASSCF method for atoms irradiated by a linearly polarized laser pulse, as detailed in Ref.~\cite{Sato2016PRA}.
Our implementation employs a spherical harmonics expansion of orbitals with the radial coordinate discretized by a finite-element discrete variable representation \cite{Rescigno:2000,McCurdy:2004,Schneider:2006,Schneider:2011}.
The computationally most costly operation is to evaluate the pair potentials [Eq.~(\ref{eq:meanfield})] contributing to the mean-field [Eq.~(\ref{eq:fock2})], for which we use a Poisson solver thereby achieving linear scaling with the number of basis functions (or equivalently, grid points) \cite{McCurdy:2004, Hochstuhl:2011, Sato2013PRA, Omiste:2017, Erik:2018}. 
A split-operator propagator is developed with an efficient implicit method for stiff derivative operators which drastically stabilizes the temporal propagation of orbitals. 
Thanks to the combination of these techniques, we can take full advantage of the TD-CASSCF method.

\subsection{Gauge Invariance}
\label{subsec:Gauge Invariance}

\index{Gauge invariance}
The TD-CASSCF method is gauge invariant. For a TD-MCSCF method to be gauge invariant, it must meet the following two requirements:
\begin{enumerate}
	\item Any LG wave function $\Psi_{\rm L}(t)$ that satisfies a given multiconfiguration ansatz Eq.~(\ref{eq:general-mcwf}) can be transformed to a VG wave function $\Psi_{\rm V}(t)$ that satisfies another multiconfiguration ansatz of the same form, and vice versa.
	\item If a LG wave function $\Psi_{\rm L}(t)$ is optimized on the basis of the TDVP expressed in the length gauge, its VG counterpart $\Psi_{\rm V}(t)$ satisfies the TDVP in the velocity gauge, and vice versa.
\end{enumerate}

To discuss the first requirement, let us denote the orbital functions calculated with a given multiconfiguration ansatz Eq.~(\ref{eq:general-mcwf}) within the length gauge by $\{\psi_p^{\rm L}({\bf r})\}$. Equation (\ref{eq:gauge-transformation}) is fulfilled if one constructs the wave function $\Psi_{\rm V}(t)$ of the same ansatz with the CI coefficients unchanged using the orbital functions $\{\psi_p^{\rm V}({\bf r})\}$ defined by,
\begin{equation}
\label{eq:orbital-gauge-transformation}
\psi_p^{\rm V}({\bf r}) = \exp \left[-i{\bf A}(t)\cdot {\bf r}\right] \psi_p^{\rm L}({\bf r}).
\end{equation}
Since this tells us that at least one of $\{\psi_p^{\rm L}({\bf r})\}$ and $\{\psi_p^{\rm V}({\bf r})\}$ is necessarily time-dependent, TD-MCSCF methods that use time-independent orbital functions such as TDCIS are, in general, not gauge invariant, i.e., the values of the observables obtained within the length gauge are not equal to those within the velocity gauge. 
This is because $\Psi_{\rm V}(t)$ does not necessarily belong to the subspace of the Hilbert space spanned by $\{\Phi_{\bf I}\}$, in which $\Psi_{\rm L}(t)$ is optimized.
It should be noticed that even if we could use an infinite number of orbitals, TDCIS would not be gauge-invariat;
it follows from Eq.~(\ref{eq:HgsVG}) that if we use time-independent orbitals and $\Psi_{\rm L}(t)$ is expressed as a single (Hartree-Fock) determinant, $\Psi_{\rm V}(t)$ involves up to $N$-tuple excitations.
(see Ref.~\cite{Sato2018AS} for a recently reported gauge-invariant formulation of TDCIS with time-dependent orbitals.)

For the second requirement, it should be noticed that the length- and velocity-gauge Hamiltonians $\hat{H}_{\rm L}(t)$ with Eq.~(\ref{eq:length-gauge}) and $\hat{H}_{\rm V}(t)$ with Eq.~(\ref{eq:velocity-gauge}), respectively, are related by \cite{Bandrauk2013JPB},
\begin{equation}
\hat{H}_{\rm V} = \hat{\mathcal{U}}\hat{H}_{\rm L}\hat{\mathcal{U}}^\dag+i\frac{d\hat{\mathcal{U}}}{dt}\hat{\mathcal{U}}^\dag.
\end{equation}
Then, using the unitarity of the gauge-transformation operator $\hat{\mathcal{U}}(t)$[Eq.~(\ref{eq:gauge-transformation-operator})], we can show that the TDVP expressions Eq.~(\ref{eq:TDVP}) in the two representations are equivalent. This guarantees that the wave function transformed via Eq.~(\ref{eq:orbital-gauge-transformation}) from the wave function satisfying the length-gauge TDVP fulfills the velocity-gauge TDVP. 
Therefore, satisfying both of the above-mentioned conditions, TD-MCSCF methods with time-varying orbital functions, including TDHF, MCTDHF, TD-CASSCF, and the time-dependent occupation-restricted multiple active-space (TD-ORMAS) \cite{Sato2015PRA} methods, are gauge invariant in general \cite{Sato2013PRA,Miyagi2014PRA,Sato2015PRA,Ishikawa2015JSTQE}

\subsection{Size Extensivity}
\label{subsec:Size Extensivity}
\index{Size Extensivity}

The TD-CASSCF method is size extensive. Size extensivity\footnote{It is not to be confused with a similar but different concept of size consistency, which, for the case of the ground-state energy, states ``if molecule AB dissociates to molecules A and B, the asymptote of molecule AB at infinite internuclear separation should be the sum of the energies of molecules A and B" \cite{Veszpremi} and ``is only defined if the two fragments are non-interacting" \cite{Jensen}.} states ``the method scales properly with the number of particles" \cite{Veszpremi} or, for the case of the ground-state energy, ``if we have $k$ number of noninteracting identical molecules, their total energy must be $k$ times the energy of one molecule" \cite{Jensen}.
Roughly speaking, it can be understood as follows.

\begin{figure}[tb]
\begin{center}
\includegraphics[width=\textwidth]{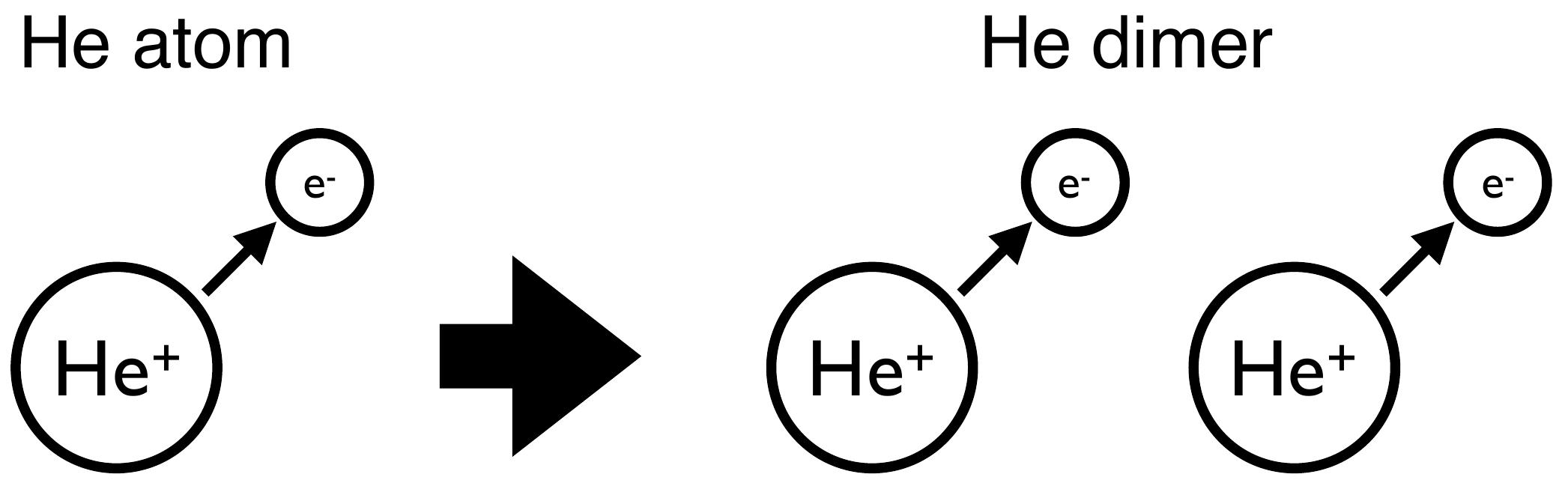} 	
\end{center} 
\caption{Schematic illustration of size extensivity explained with a He atom and dimer (see text).}
\label{fig:size-extensivity}
\end{figure}

Let us consider that we simulate photoionization of a He atom for such a laser parameter that He is substantially singly ionized but that double ionization is negligible. 
Then, what will happen if we simulate photoionization of a He dimer by the identical laser pulse, in which the two He atoms are sufficiently far apart from each other but the dipole approximation is still valid? 
Physically, we would expect substantial single ionization of each atom, resulting in double ionization in total (Fig.~\ref{fig:size-extensivity}).

This seemingly obvious requirement is, in general, {\it not} met by TD-MCSCF methods with truncated expansion such as TDCIS and TD-ORMAS.
On the other hand, TD-CASSCF as well as MCTDHF and TDHF fulfills size extensivity.

\section{Initial-State Preparation and Simulation Boundary}
\label{sec:Initial-State Preparation and Simulation Boundary}

In {\it ab initio} simulation study of multielectron dynamics, we usually need to (i) prepare the initial state, (ii) propagate the wave function in time (Sec.~\ref{sec:TD-CASSCF}), (iii) absorb electrons that leave the calculation region, and (iv) read out physically relevant information from the wave function (Sec~\ref{sec:Numerical examples}). 
Let us discuss (i) and (iii) in this Section.

\subsection{Imaginary-Time Propagation}
\index{Imaginary-time propagation}

While the initial state can also be obtained by a separate time-independent calculation of the ground state, a convenient alternative is imaginary-time propagation (or relaxation) \cite{Flocard1978PRC}.
The solution of the field-free TDSE can be expressed as,
\begin{equation}
	\Psi (t) = \sum_{\alpha=0}^\infty c_\alpha \Xi_\alpha e^{-iE_\alpha t}= e^{-iE_0 t} \left(c_0 \Xi_0  + \sum_{\alpha=1}^\infty c_\alpha \Xi_\alpha e^{-i(E_\alpha-E_0) t}\right),
\end{equation}
with eigenstates $\Xi_\alpha$, of which $\Xi_0$ is the ground state, and energy eigenvalues $E_\alpha$.
By substituting {\it imaginary} time $t = -is$ with $s$ being a real number, we obtain,
\begin{equation}
	\Psi (-is) e^{E_0 s} =  c_0 \Xi_0  + \sum_{\alpha=1}^\infty c_\alpha \Xi_\alpha e^{-(E_\alpha-E_0) s} \xrightarrow[s\to \infty]{} c_0 \Xi_0,
\end{equation}
since $E_\alpha-E_0>0$ ($\alpha\ge 1$). Thus, we can obtain the ground state by integrating the field-free EOMs in imaginary time and renormalizing the wave function after every several time steps.
The imaginary-time propagation is used for the results presented in this Chapter.

\subsection{Absorption Boundary}

Since ionization is essential in the ultrafast intense laser science, it is one of the major issues how to treat electrons that leave the calculation region and suppress unphysical reflections.
We use either mask function or infinite-range exterior complex scaling in our numerical implementations.

\subsubsection{Mask Function and Complex Absorbing Potential (CAP)}

One common method is to multiply orbital functions outside a given radius (mask radius) $R_0$ by a function that decreases from unity and vanishes at the simulation box boundary after each time step\index{Mask function} \cite{Krause_1992}. 
Typical forms of the mask function include $\cos^{1/4}$ and $\cos^{1/8}$.

Another method is to add a complex absorbing potential (CAP), e.g., of the form
\begin{equation}
	-i\eta W(r) = -i \eta (r-R_0)^2,
\end{equation}
where $\eta$ denotes a CAP strength, to the Hamiltonian outside a given radius $R_0$. In the context of {\it ab initio} simulations of strong-field processes, CAP has been used in combination with TDCIS \cite{Greenman_2010}.

\subsubsection{Exterior Complex Scaling (ECS)}

\begin{figure}[tb]
\sidecaption
\includegraphics[width=0.5\textwidth]{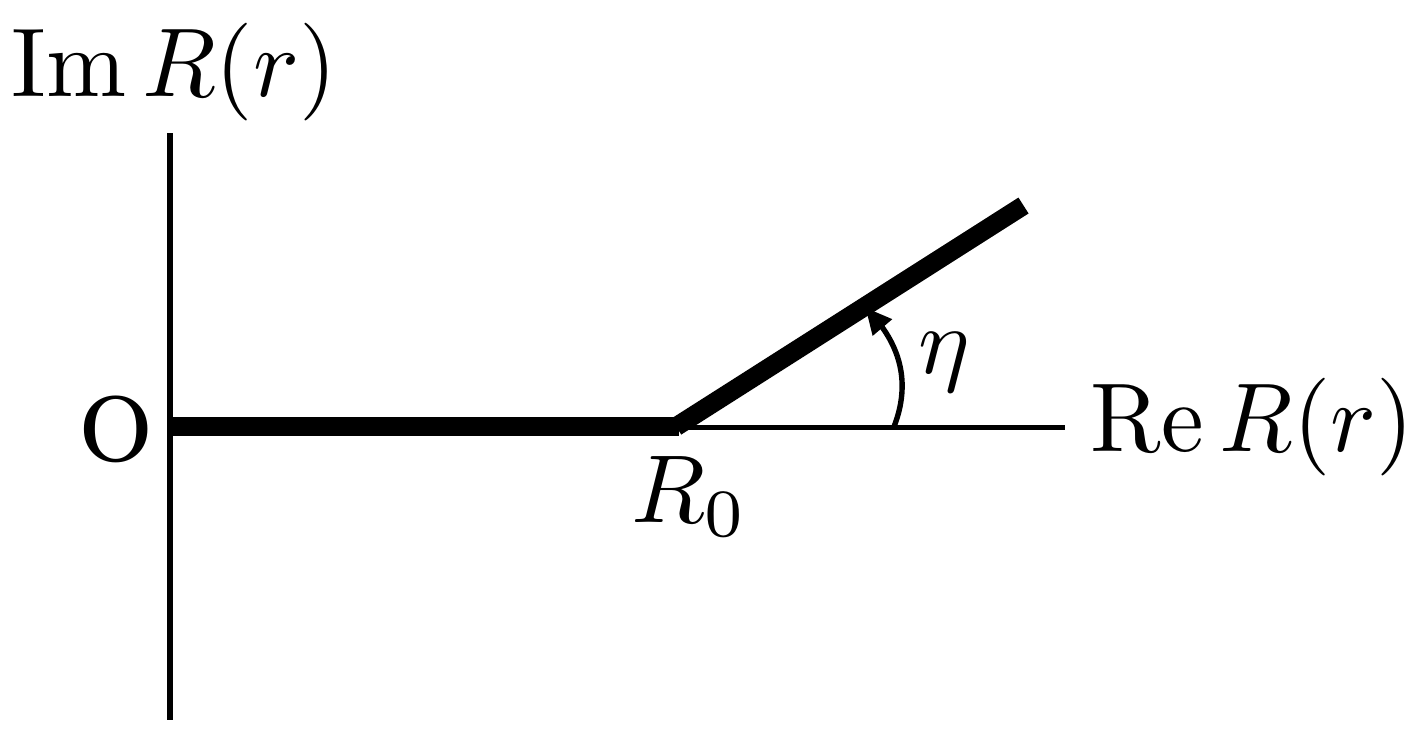}
\caption{Schematic illustration of radial exterior complex scaling contour $R(r)$ with scaling radius $R_0$ and scaling angle $\eta$.}
\label{fig:ECScontour}
\end{figure}

Exterior complex scaling\index{Exterior complex scaling (ECS)}\cite{McCurdy_1991} analytically continues the wave function outside a given scaling radius $R_0$ into the complex plane as, for the case of the polar coordinate (Fig.~\ref{fig:ECScontour}),
\begin{equation}
	\label{eq:rchange}
	r \rightarrow R(r) = 
	\begin{cases}
		r & (r < R_0) \\
		R_0 + (r-R_0) e^{\lambda + i \eta} & (r > R_0),
	\end{cases}
\end{equation}
where $\lambda$ and scaling angle ${\eta}$ are real numbers. Then, the orbital function is transformed via ECS operator $U_{\eta R_0}$ as,
\begin{equation}\label{eq:Ueta}
	(U_{\eta R_0} \psi) (\vec{r}): = 
	\begin{cases}
		\psi( \vec{r} ) & (r < R_0) \\
		e^{\frac{\lambda + i \eta}{2}} \dfrac{R(r)}{r} \psi( \vec{R}(r)) & (r > R_0),
	\end{cases}
\end{equation}
where $\vec{R}(r) = \frac{R(r)}{r}\vec{r}$.
In actual simulations, we numerically store $(U_{\eta R_0} \psi) (\vec{r})$ instead of $\psi( \vec{r} )$ in the scaled region $r > R_0$.
We can understand why this works as an absorbing boundary by considering a spherical wave $e^{ikr}/r$. At $r>R_0$ it becomes $e^{\frac{\lambda + i \eta}{2}}e^{ik[R_0 + (r-R_0) e^{\lambda} \cos\eta]-k(r-R_0)\sin\eta}/r$, which exponentially diminishes as $\sim e^{-kr\sin\eta}/r$ at large distance.
It should be noticed that ECS modifies neither the wave function nor the system Hamiltonian.

While ECS is usually applied on a finite discretization range, one can infinitely extend the scaled region, thus moving the simulation box boundary to infinity, while using a finite number of exponentially damped basis functions \cite{Scrinzi_2010}. This method, called infinite-range ECS (irECS)\index{Infinite-range exterior complex scaling (irECS)}, significantly improves the accuracy and efficiency over standard ECS with a considerably smaller number of basis functions. It also has a conceptual advantage of being able to simulate the entire space.

While irECS has originally been formulated for a single-electron system and found only limited use for strongly-driven multielectron systems, we have applied it to our TD-CASSCF numerical implementation, as detailed in Ref.~\cite{Orimo2018PRA}.
We set $\lambda=0$ and introduce Gauss-Laguerre-Radau quadrature points \cite{Gautschi_2000,Weinmueller_2017} to construct discrete-variable-representation basis functions in the last finite element extending to infinity.
An essential difference from a single-electron case is the presence of interelectronic Coulomb interaction via mean-field operator Eq.~(\ref{eq:fock2}). Its evaluation as well as that of $\hat{Q}$ requires $\hat{U}_{(-\eta) R_0} \Ket{\psi_p}$, which is not available in the scaled region.
Since the scaled region is far from the origin, it is reasonable to assume that the scaled part of the orbital functions hardly affects the electron dynamics close to the nucleus and that the interaction between electrons residing in the scaled region is negligible.
Thus, we neglect $\hat{U}_{(-\eta) R_0} \Ket{\psi_p}$ in the scaled region wherever their information is necessary. This treatment roughly corresponds to the neglect of the Coulomb force acting on electrons from scaled-region electrons ($r > R_0$). On the other hand, the Coulomb force acting on scaled-region electrons from unscaled-region electrons ($r < R_0$) is not neglected. Hence, the effect of the ionic Coulomb potential is properly taken into account in the dynamics of departing electrons.

\begin{figure}[tb]
\includegraphics[width=\textwidth]{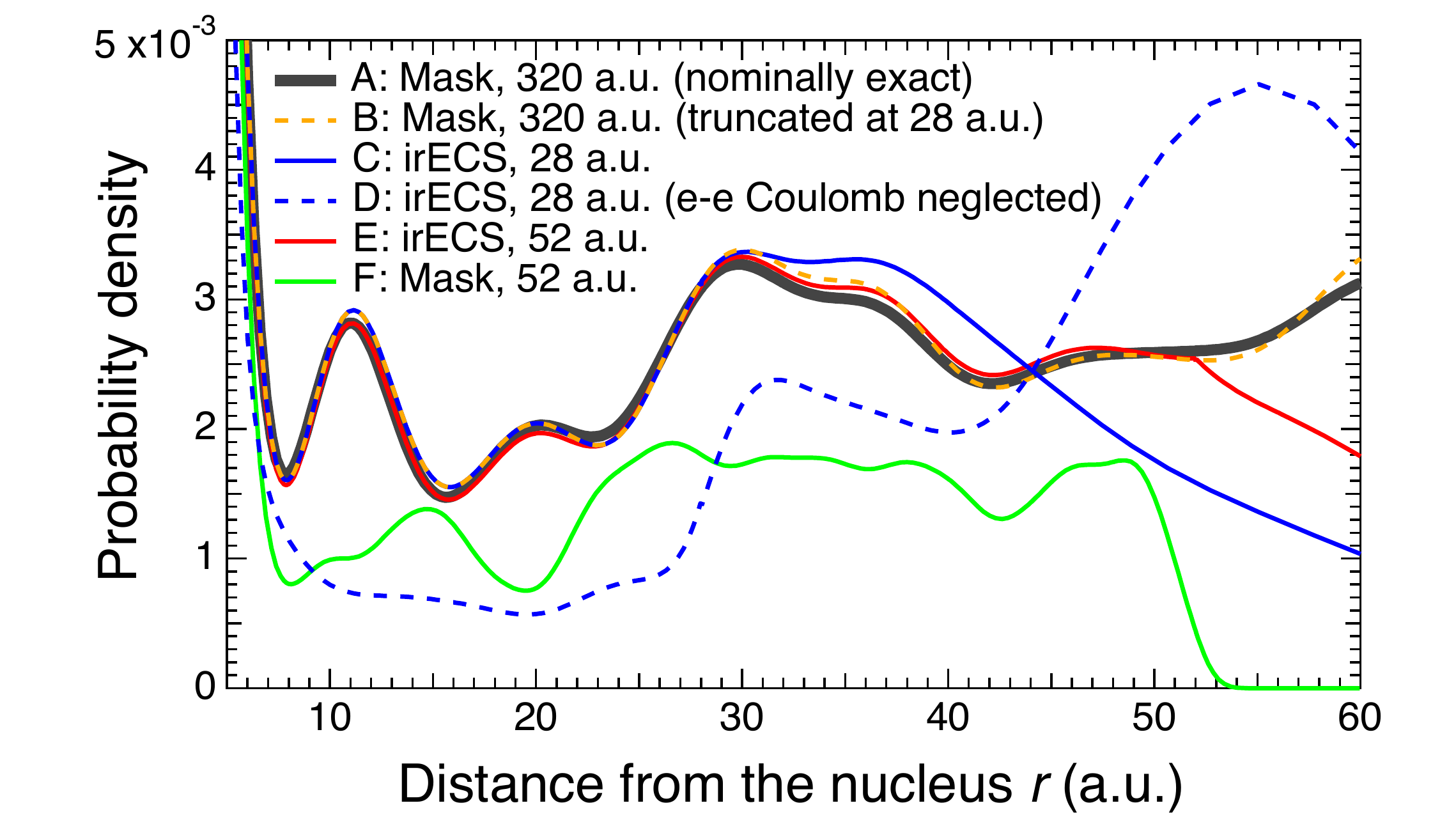}
\caption{Electron radial distribution function $\rho (r)$ after the laser pulse for the case of Be exposed to a laser pulse with 800 nm wavelength and $3.0 \times 10^{14} \text{ W/m$^2$}$ peak intensity, calculated with different absorbing boundaries listed in Table \ref{tab:detailBe}.}
\label{fig:radBe}
\end{figure}

Figure \ref{fig:radBe} compares the electron radial distribution functions after the pulse for the case of a Be atom exposed to a laser pulse with 800 nm wavelength and $3.0 \times 10^{14} \text{ W/m$^2$}$ peak intensity, calculated with different absorbing boundaries listed in Table \ref{tab:detailBe}.
The pulse has a $\sin^2$ envelope with a foot-to-foot pulse width of five cycles. We use $(n_{fc},n_{dc},n_{a}) = (1,0,4)$.
The result of condition A, with $R_0=320$ a.u. much larger than the quiver radius 28.5 a.u., is converged and can be considered to be numerically exact. We can see that the irECS delivers much better results (C and E) inside $R_0$ than the mask function (F).
It is remarkable that the irECS works well even with the scaling radius ($R_0=28$ a.u.) comparable with the quiver radius.
The result of the simulation (condition D) similar to C but neglecting also the interelectronic Coulomb force from the unscaled (inner) to the scaled (outer) region is plotted with a blue dashed curve in Fig.~\ref{fig:radBe}.
We find a large discrepancy from the exact result (A).
This indicates that proper account of the Coulomb force acting on scaled-region electrons from unscaled-region ones is crucial for accurate simulations, even though the total momentum of the system is not conserved due to imbalance in counting the interelectronic Coulomb interactions.


\begin{table}[tb]
	\caption{Absorbing boundaries tested for Be.}
		\begin{tabular}{p{1cm}p{2cm}p{1cm}p{1cm}p{1cm}p{1cm}p{4cm}}
			\hline
			& Absorber & $R_0$ & $n_\text{ua}$ & $L_{\text{a}}$ & $n_\text{a}$ & Remark\\
			\svhline
			A & mask & 320 & 1600 & 80 & 400 & nominally exact\\
			B & mask & 320 & 1600 & 80 & 400 & truncated at $28$ a.u. (see text in Sebsec.~\ref{subsubsec:after-absorption})\\
			C & irECS & 28 & 140 & $ \infty $ & 40 \\
			D & irECS & 28 & 140 & $ \infty $ & 40 & unscaled-to-scaled Coulomb neglected \\
			E & irECS & 52 & 260 & $ \infty $ & 40 \\
			F & mask & 52 & 260 & $ 8 $ & 40 \\
			\hline
		\end{tabular}
	\label{tab:detailBe}%
\end{table} 

\subsubsection{Which part of the total wave function is propagated after one or more electrons are absorbed at the simulation boundary?}
\label{subsubsec:after-absorption}

\begin{figure}[tb]
\sidecaption
\includegraphics[width=0.5\textwidth]{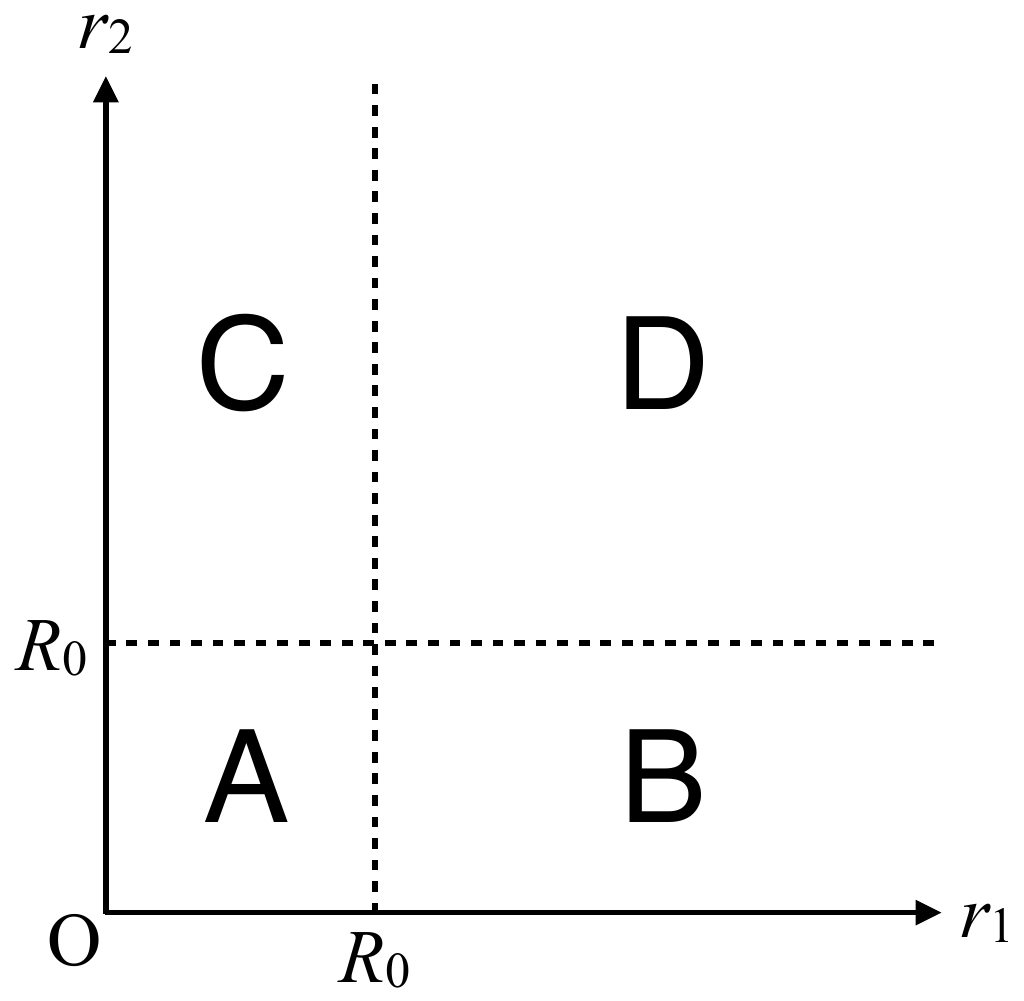}
%
%
\caption{Four regions of the $(r_1,r_2)$ space of the two electrons in He.}
\label{fig:He-regions}       
\end{figure}

Let us specifically consider a He atom, which is a two-electron system. 
The $(r_1,r_2)$ space can be divided into four regions, as shown in Fig.~\ref{fig:He-regions}, A: $r_1<R_0, r_2<R_0$, B: $r_1 > R_0, r_2 < R_0$, C: $r_1 < R_0, r_2 > R_0$, D: $r_1 > R_0, r_2 > R_0$.

For the case of direct numerical simulation of the two-electron TDSE, e.g., by the time-dependent close-coupling method \cite{Colgan2001JPB, Parker2001, ATDI2005, Feist2009PRL}, the wave function only in region A is stored and propagated. Hence, once one electron is absorbed, the dynamics of the other electron is no longer followed even if it is still inside the absorption radius $R_0$, and, as a consequence, the transition from B or C to D cannot be traced.

In great contrast, not only the two electrons in region A but also the inner electron in regions B and C is simulated in the TD-CASSCF, MCTDHF, and TDHF simulations.
In order to understand this prominent feature, we decompose the exact orbital $\ket{\psi_p}$, which would be obtained if we used an infinitely large simulation box, into the part numerically stored and propagated during actual simulation $\ket{\phi_p}$ and the remaining, i.e., absorbed and lost part $\ket{\chi_p}$:
\begin{equation}
	\ket{\psi_p} = \ket{\phi_p} + \ket{\chi_p}.
\end{equation}
Roughly speaking, $\ket{\phi_p}$ is the part at $r<R_0$ and $\ket{\chi_p}$ at $r>R_0$. The TD-CASSCF equations of motion are derived on the assumption that $\{\ket{\psi_p}\}$ is orthonormal. On the other hand, $\{\ket{\phi_p}\}$ is {\it not} orthonormal in general, and its norm decreases. By good use of absorption boundary, $\ket{\phi_p}$ expectedly reproduces $\ket{\psi_p}$ within $R_0$.
In region B, the two-electron wave function is generally expressed as,
\begin{equation}
	\label{eq:He-expansion}
	\Psi ({\bf r}_1,{\bf r}_2) = \sum_{p,q} C_{pq}\chi_p ({\bf r}_1) \phi_q ({\bf r}_2),
\end{equation}
with an expansion coefficient $C_{pq}$. Here we neglect the spin part for simplicity. As Eq.~(\ref{eq:He-expansion}) suggests, even after electron 1 is absorbed, the dynamics of electron 2, still entangled with electron 1, continues to be simulated, though approximately, as long as it stays inside the absorption radius.

In Fig. \ref{fig:radBe} we have seen that the irECS works much better than the mask function. Nevertheless, the irECS results (conditions C and E in Table \ref{tab:detailBe}) still deviate slightly from the nominally exact solution (condition A).
In the present case, Be is nearly totally ionized, and double ionization amounts to 50 \%, as we will see below in Fig. \ref{fig:ipxBe}. Hence, the deviation may be due to the neglect of the Coulomb interaction in and from the scaled region and/or the loss of information on the wave function in the scaled region.

In order to reveal the effect of the latter, we have performed a simulation (condition B in Table \ref{tab:detailBe}) with a sufficiently large domain as condition A but by truncating the interelectronic Coulomb interaction at $r=28\,{\rm a.u.}$ as in the irECS.
The result is plotted in Fig.~\ref{fig:radBe}.
The ``truncated" result (B) slightly deviates from the exact one (A) but overlaps with the irECS result (C) at $r<28$ a.u., which indicates that the difference between the exact and irECS results in Fig.~\ref{fig:radBe} originates from the neglect of the Coulomb interaction in and from the scaled region, not from the loss of information.

One may be surprised that the loss of information on orbital functions at the absorption boundary hardly affects simulation results within the absorption radius.
It should be noticed that, even if the explicit form of $\ket{\chi_p}$ is unknown, some information on them is still available. At least, we can tell,
\begin{equation}
		\langle \phi_p | \chi_q \rangle = 0	, \qquad \langle \chi_p | \chi_q \rangle = \delta_{pq} - \langle \phi_p | \phi_q \rangle,
\end{equation}
from the orthonormality of $\{\ket{\psi_p}\}$. 
This not only helps accurate simulations but also allows to extract useful information such as ionization yields and charge-state-resolved observables, as discussed in the next section. 

\section{Numerical examples}
\label{sec:Numerical examples}

In this Section, we present how to extract physical observables from the wave function and numerical results obtained with TD-CASSCF and TDHF simulations.

\subsection{Ionization Yield}
\label{subsec:Ionization Yield}

One might consider that the ionization yield for charge state $n$ could be obtained through the integration over the population of all possible $n$-electron continuum states (note that $n$ denotes the ionic charge state in this Section). 
Unfortunately, however, direct application of this naive idea would encounter difficulties. 
First, it is not trivial (even more difficult than TD-CASSCF itself) to prepare $n$-electron continuum wave functions. The ionic core is not necessarily in the ground state and may also be excited. 
Second, we have to keep the entire wave function within the simulation box, without being absorbed. The computational cost would be prohibitive. 
Third, as discussed in Subsec.~\ref{subsec:Gauge Transformation}, the population of each field-free stationary state is not gauge invariant during the pulse.

Instead, we define ionization in terms of the spatial positions of electrons and introduce ionization probability $P_n$ as a probability
to find $n$ electrons in the outer region $|{\bf r}| >
R_{\rm ion}$ and the remaining $N - n$ 
electrons in the inner region $|{\bf r}| < R_{\rm ion}$,
with a given distance $R_{\rm ion}$ from the origin,
\begin{eqnarray}
\label{eq:ionp} 
P_n &\equiv& \binom{N}{n} 
\int_> dx_1 \cdot\cdot
\int_> dx_n
\int_< dx_{n+1} \cdot\cdot
\int_< dx_N \,
\left| \Psi (x_1,\cdot\cdot,x_N) \right|^2,
\end{eqnarray} 
where $\int_<$ and $\int_>$ denote integrations over a spatial-spin
variable $x = \{{\bf r}, \sigma\}$ with the spatial part restricted to the domains
$|{\bf r}| < R_{\rm ion}$, and $|{\bf r}| > R_{\rm ion}$, respectively.
${P_n}$ satisfies $\sum_{n=0}^N P_n = 1$.
This spatial-domain-based ionization probability has an advantage of being gauge invariant. Moreover, it is consistent with our usual perception of ionization as a spatial separation of electron from the parent ion, such as ejection from the surface and arrival of electron at a detector.

If we introduce,
\begin{eqnarray}
\label{eq:ionp-aux} 
T_n &\equiv& \binom{N}{n} 
\int dx_1 \cdot\cdot
\int dx_n
\int_< dx_{n+1} \cdot\cdot
\int_< dx_N \,
\left| \Psi (x_1,\cdot\cdot,x_N) \right|^2,
\end{eqnarray} 
it is related to $P_n$ as,
\begin{eqnarray}
\label{eq:ionp-t}
P_n = \sum_{k=0}^n \binom{N-n+k}{k}(-1)^k T_{n-k},
\end{eqnarray}
due to the orthonormality of orbitals with full-space integration \cite{Sato2013PRA} (see also Subsec.~\ref{subsubsec:after-absorption}), allowing to calculate the ionization probability only from the information of orbitals inside the radius $R_{\rm ion}$ and CI coefficients.
By adopting the multiconfiguration expansion Eq.~(\ref{eq:general-mcwf}), and making use of
the orthonormality of spin-orbitals in the full-space integration, we
have 
\begin{eqnarray}
\label{eq:iont}
T_n = \sum_{IJ}^{\sf P} C^*_I C_J D^{(n)}_{IJ},
\end{eqnarray}
where,
\begin{eqnarray} 
\label{eq:iond}
D^{(0)}_{IJ} &=& \sum_{ij}^N
\det\left(S^<_{IJ}\right), \nonumber \\
D^{(1)}_{IJ} &=& \sum_{ij}^N \epsilon^{IJ}_{ij} 
\det\left(S^<_{IJ}[i:j]\right), \nonumber \\
D^{(2)}_{IJ} &=& \sum_{i>j}^N \sum_{k>l}^N \epsilon^{IJ}_{ik}
 \epsilon^{IJ}_{jl}
\det\left(S^<_{IJ}[ij:kl]\right),
	       \end{eqnarray}
etc, and  $S^<_{IJ}$ is an $N\times N$
matrix with its $\{ij\}$ element being the inner-region overlap integral,
\begin{eqnarray}
\label{eq:overlap}
(S^<_{IJ})_{ij} = \int_< dx \phi^*_{p(i,I)}(x) \phi_{q(j,J)}(x) \equiv
\langle \phi_p | \phi_q \rangle_<,
\end{eqnarray}
where $\phi_{p(i,I)}$ is the $i$-th (in a predefined order) spin orbital in
the determinant $I$. $S^<_{IJ}[ij\cdot\cdot:kl\cdot\cdot]$ is the
submatrix of $S^<_{IJ}$ obtained after removing rows $i,j,\cdot\cdot$ and columns
$k,l,\cdot\cdot$ from the latter, and,
\begin{eqnarray}
\label{eq:epsilon}
\epsilon^{IJ}_{ij} = \delta^{p(i,I)}_{q(j,J)} (-1)^{i+j}.
\end{eqnarray}
The matrix $S^<_{IJ}$ and its submatrices are block-diagonal due to
the spin-orthonormality, so that, e.g., $\det\left(S^<_{IJ}\right) = 
\det\left(S^<_{I^\alpha J^\alpha}\right) 
\det\left(S^<_{I^\beta J^\beta}\right)$, where $I^\sigma$ is the
$\sigma$-spin part of the determinant $I$.

\begin{figure}[tb]
	\begin{center}
		\includegraphics[width=0.75\textwidth]{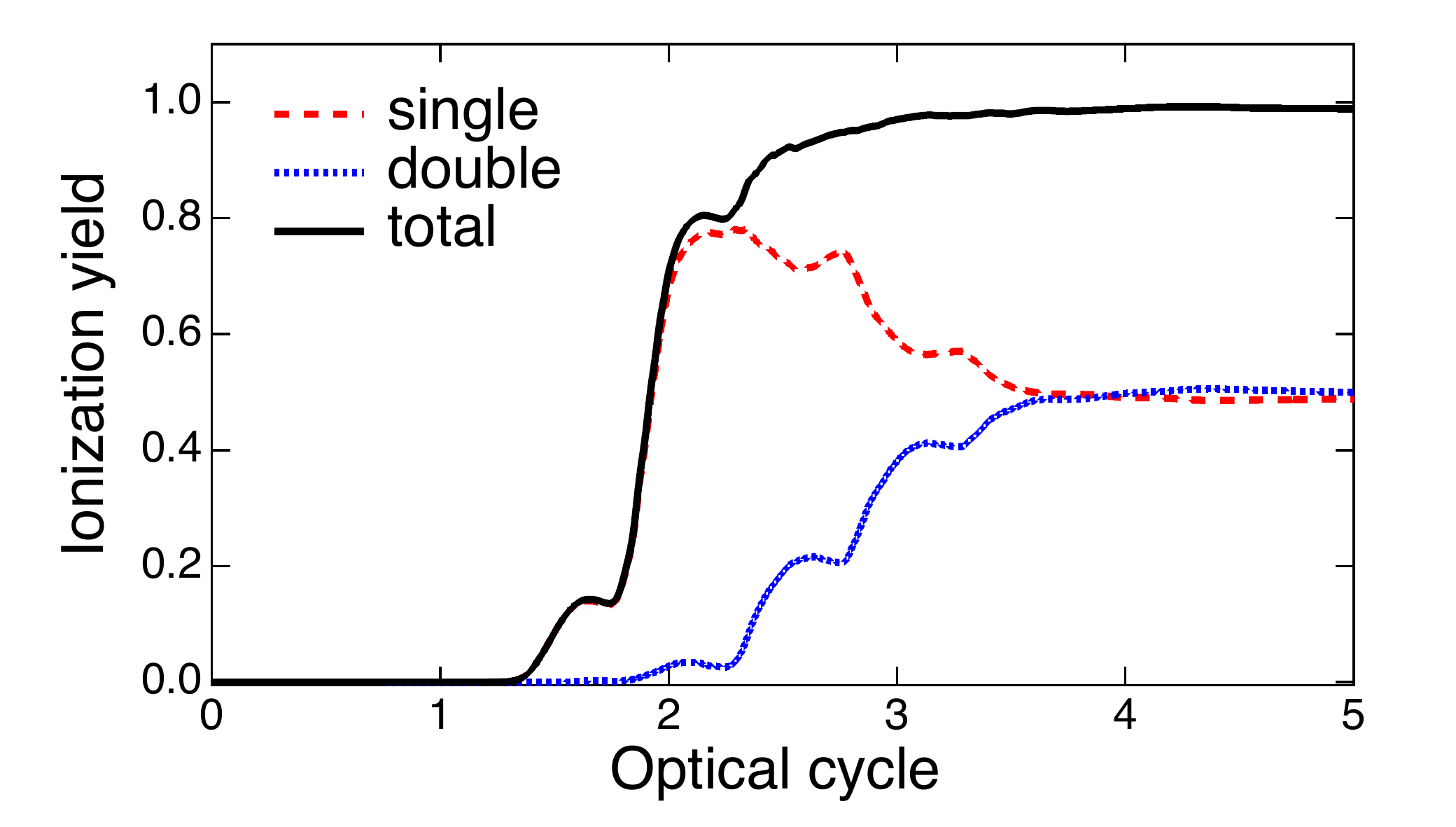}
	\end{center}
	\caption{Time evolution of spatial-domain-based single, double, and total ionization probabilities of Be exposed to a laser pulse with 800 nm wavelength and $3.0 \times 10^{14} \text{ W/m$^2$}$ peak intensity (the same as for Fig.~\ref{fig:radBe}). $R_{\rm ion} = 20\,{\rm a.u.}$ is used.}
	\label{fig:ipxBe}
\end{figure}

In Fig.~\ref{fig:ipxBe}, we show the temporal evolution of thus calculated single, double, and total ionization yields of Be for the same pulse and orbital subspace decomposition as for Fig.~\ref{fig:radBe}.
As an absorption boundary, we have used irECS with $R_0 = 40\,{\rm a.u.}$. $R_{\rm ion}$ is set to be $20\,{\rm a.u.}$.
We can see step-like evolution every half cycle typical of tunneling ionization. 
After the pulse, there is practically no neutral species left, and the double ionization yield is $\sim 50\%$.
It is remarkable that the neglect of the Coulomb interaction in and from the scaled region is a good approximation and that irECS works excellently even under such massive double ionization.

\begin{figure}[tb]
	\begin{center}
		\includegraphics[width=\textwidth]{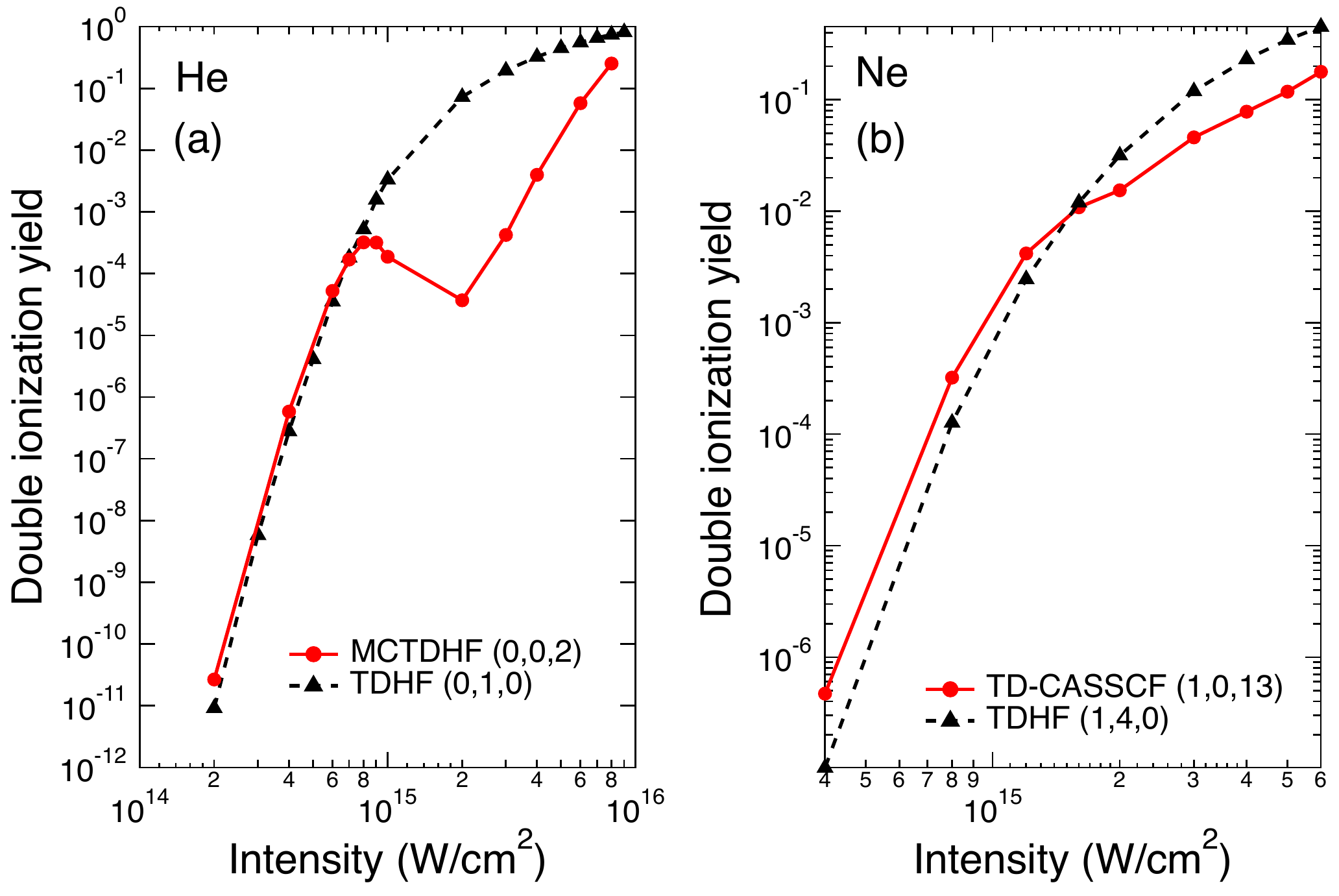}
	\end{center}
	\caption{Double ionization yields of (a) He and (b) Ne as a function of intensity of a laser pulse with a wavelength of 800 nm, calculated by the TD-CASSCF and TDHF methods.}
	\label{fig:NSDI}
\end{figure}

Figure \ref{fig:NSDI} presents the intensity dependence of the double ionization yields of He and Ne irradiated by a laser pulse whose wavelength is 800 nm. 
Although the results are not converged with respect to the number of orbitals yet, we can already clearly see knee structures in the TD-CASSCF results, but not in the TDHF ones.
Thus, the TD-CASSCF method can well reproduce non-sequential double ionization \cite{Walker1994,Larochelle1998}, a representative strong-field phenomenon that witnesses electron correlation. 

\subsection{Charge-State-Resolved Electron Density Distribution}

The usual electron density distribution,
\begin{equation}
\label{eq:EDD}
	\rho ({\bf r}) \equiv N \sum_{\sigma}
\int dx_2 \cdots
\int dx_N \,
\left| \Psi (x,x_2,\cdots,x_N) \right|^2,
\end{equation}
contains all the different charge states: neutral, singly ionized, doubly ionized, etc.
To discuss charge migration following attosecond photoionization, it will be useful to analyze, e.g., a hole distribution inside the cation.
Hence, we introduce a charge-state-resolved electron density distribution $\rho^{(n)} ({\bf r})$ as a probability to find an electron at ${\bf r}$ on condition that $n$ out of the other $N-1$ electrons are at $|{\bf r}| > R_{\rm ion}$ and $N-1-n$ at $|{\bf r}| < R_{\rm ion}$,
\begin{equation}
\label{eq:EDD-chg}
	\rho^{(n)} ({\bf r}) \equiv N\binom{N-1}{n} \sum_{\sigma}
\int_> dx_2 \cdots
\int_> dx_{n+1}
\int_< dx_{n+2} \cdots
\int_< dx_N \,
\left| \Psi (x,x_2,\cdots,x_N) \right|^2.
\end{equation}
Note that the electron density distribution in the neutral species is consistently given by,
\begin{equation}
\label{eq:EDD-neutral}
	\rho^{(0)} ({\bf r}) \equiv N \sum_{\sigma}
\int_< dx_{2} \cdots
\int_< dx_N \,
\left| \Psi (x,x_2,\cdots,x_N) \right|^2.
\end{equation}
Here, again, we have used domain-based definition of ionization.

$\rho^{(n)} ({\bf r})$ can be expressed in terms of orbitals and CI coefficients as well as $S_{IJ}^<$ introduced in the previous Subsection. For example, the electron density distribution of the cation can be calculated by,
\begin{equation}
	\rho^{(1)}({\bf r}) = \sum_{IJ}^\Pi C^*_I C_J \sum_{i,j}^{N} \phi^*_{p(i,I)}(x) \phi_{q(j,J)}(x) (-1)^{i+j} \left( \sum_{k=1}^N \epsilon_{ik}\epsilon_{jk}S_{IJ}^<[ik;jk]-(N-1)S_{IJ}^<[i;j] \right).
\end{equation}

\begin{figure}[tb]
\sidecaption
\includegraphics[width=\textwidth]{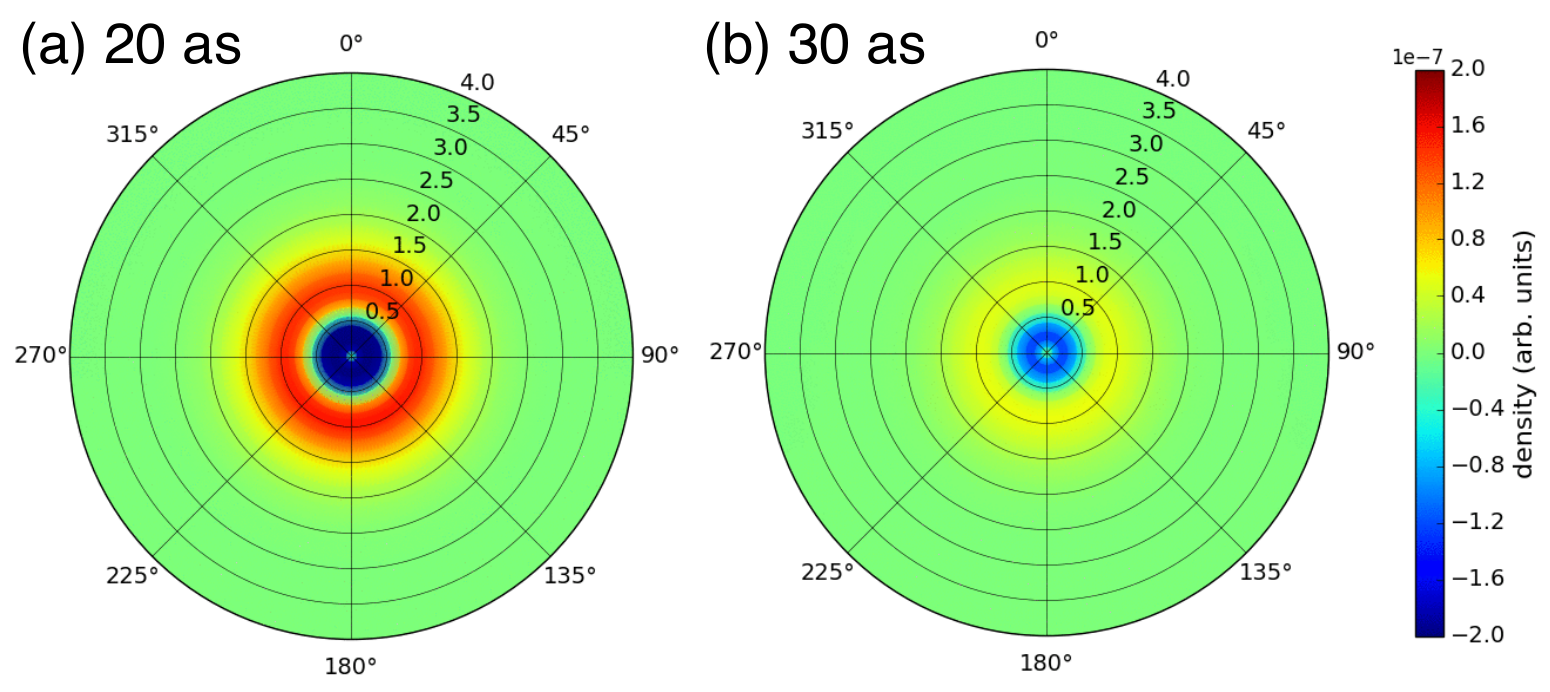}   
%
%
\caption{Snap shots of the electron density distribution in ${\rm Be}^+$ produced by photoionization of Be by XUV pulses with a photon energy of 150 eV, a peak intensity of $10^{13}\,{\rm W/cm}^2$, and a FWHM pulse width of (a) 20 as and (b) 30 as. The results of TDHF simulations.}
\label{fig:Be-charge-oscillation}       
\end{figure}

In Fig.~\ref{fig:Be-charge-oscillation}, we show snap shots of the electron density distribution in ${\rm Be}^+$ produced by photoionization of Be by XUV pulses with a photon energy of 150 eV and a FWHM pulse width of 20 and 30 as.
The process is simulated with the TDHF method.
An isotropic charge density is formed by the superposition of $(1s)^{-1}$ and $(2s)^{-1}$ holes and oscillates with a period of ca.~35 as, consistent with the energy difference ($\sim 120$ eV). We also see that its amplitude is larger for the 20 as pulse width than for 30 as, reflecting the wider spectrum of photon energy.
The present charge-state-resolved density can also be used to decompose physical observables to contributions from species of different ionic charges, e.g, charge-state-resolved HHG spectra \cite{Tikhomirov2017PRL}. 

\subsection{Ehrenfest Expression for Dipole Acceleration and High-Harmonic Spectrum}
\label{subsec:Dipole Acceleration}

Harmonic spectrum is usually extracted by Fourier transforming the dipole moment,
\begin{eqnarray}\label{eq:ehrenfest_z}
\langle z \rangle (t)  = \langle\Psi|z|\Psi\rangle,
\end{eqnarray}
or the dipole acceleration\index{Dipole acceleration} $\langle a\rangle(t)$,
\begin{eqnarray}\label{eq:a_td}
\langle a\rangle(t) = \frac{d^2}{dt^2}\langle z \rangle (t).
\end{eqnarray}
As known as the Ehrenfest theorem, one can show, from the TDSE Eq.~(\ref{eq:TDSE}), that,
\begin{eqnarray}\label{eq:ehrenfest_acc}
\langle a\rangle(t) = -
\langle\Psi|\left(
\frac{\partial \hat{V}_0}{\partial z}+
\frac{\partial \hat{V}_{\rm ext}}{\partial z}
\right)|\Psi\rangle.
\end{eqnarray}
The right hand side of this equation is the expectation value of the force acting on the electrons from the nuclei and laser electric field.
Equation~(\ref{eq:ehrenfest_acc}), with smaller numerical noise than in Eq.~(\ref{eq:ehrenfest_z}), is widely used in combination with TDSE simulations within the single-active-electron (SAE) approximation, with $\hat{V}_0$ replaced by the effective potential.

The equivalence of Eqs.~(\ref{eq:a_td}) and (\ref{eq:ehrenfest_acc}) holds also for the TD-CASSCF methods with all the orbitals time-varying \cite{Sato2016PRA}, and, hence, the Ehrenfest expression Eq.~(\ref{eq:ehrenfest_acc}) can be safely used. 
However, the use of frozen-core orbitals requires a special care.
We have shown that, in the latter case, the following expression should be used instead of Eq.~(\ref{eq:ehrenfest_acc}) \cite{Sato2016PRA}:
\begin{eqnarray}\label{eq:ehrenfest_fc_acc}
\langle a \rangle_\textrm{fc} (t) = -
\langle\Psi|\left(
\frac{\partial \hat{V}_0}{\partial z}+
\frac{\partial \hat{V}_{\rm ext}}{\partial z}
\right)|\Psi\rangle +
\Delta(\dot{p}_z).
\end{eqnarray}
In the length gauge and if we neglect the indistinguishability between core and active electrons, the additional term $\Delta(\dot{p}_z)$ can be approximated as \cite{Sato2016PRA},
\begin{eqnarray}\label{eq:gbf_fc_approx_p}
\Delta(\dot{p}_z) &\approx& 
 \langle\Phi_\textrm{fc}|
 \frac{\partial \hat{V}_0}{\partial z} +
 \frac{\partial \hat{V}_{\rm ext}}{\partial z} +
 \frac{\partial \hat{V}_a}{\partial z}
 |\Phi_\textrm{fc}\rangle,
\end{eqnarray}
where,
\begin{eqnarray}\label{eq:coulomb_act}
V_a({\bf r}) = \int d{\bf r}^\prime \frac{\rho_a({\bf r}^\prime)}{|{\bf r}-{\bf r}^\prime|},
\end{eqnarray}
with $\rho_a$ being the density of active electrons.

The meaning of Eq.~(\ref{eq:ehrenfest_fc_acc}) can be interpreted as follows: 
The Ehrenfest theorem states that the dipole acceleration is given by the expectation value of the total force on the electronic system, made up of the laser electric force acting on the active, $f_{\rm la}$, and core electrons, $f_{\rm lc}$, the nuclear Coulomb force on the active, $f_{\rm na}$, and core electrons, $f_{\rm nc}$, and the interelectronic forces from the active electrons on the core, $f_{\rm ac}$, and vice versa, $f_{\rm ca}$.
Then, we obtain the total force,
\begin{equation}\label{eq:force_ehrenfest}
f = (f_{\rm na} + f_{\rm nc}) + (f_{\rm la} + f_{\rm lc}) + (f_{\rm ac} + f_{\rm ca}) = (f_{\rm na} + f_{\rm nc}) + (f_{\rm la} + f_{\rm lc}),	
\end{equation}
where we have used the action-reaction law $f_{\rm ac}=-f_{\rm ca}$ in the second equality.
We can see correspondence of this expression to Eq.~(\ref{eq:ehrenfest_acc}).
However, if the core orbitals are frozen, we have to take account of an additional ``binding force'' $f_{\rm b}$ to fix the frozen core, which is inherent in the variational procedure to derive the EOMs.
Since the binding force $f_{\rm b}$ cancels the forces acting on frozen-core electrons from the nuclei, laser field, and active electrons, it is given by,
\begin{eqnarray} \label{eq:force_ehrenfest3}
f_{\rm b} = - f_{\rm nc} - f_{\rm lc} - f_{\rm ac}.
\end{eqnarray}
Consequently, the effective force in the presence of frozen core becomes, 
\begin{eqnarray} \label{eq:force_fc}
f_{\rm eff} = f + f_{\rm b} = (f_{\rm na} + f_{\rm ca}) + f_{\rm la}.
\end{eqnarray}
The comparison between Eqs.~(\ref{eq:gbf_fc_approx_p}) and (\ref{eq:force_ehrenfest3}) tells us that the additional term $\Delta(\dot{p}_z)$ in the former represents the binding force $f_{\rm b}$, and Eq.~(\ref{eq:ehrenfest_fc_acc}) is a quantum-mechanical expression of the effective force [Eq.~(\ref{eq:force_fc})].

\begin{figure}[tb]
	\begin{center}
		\includegraphics[width=0.75\textwidth]{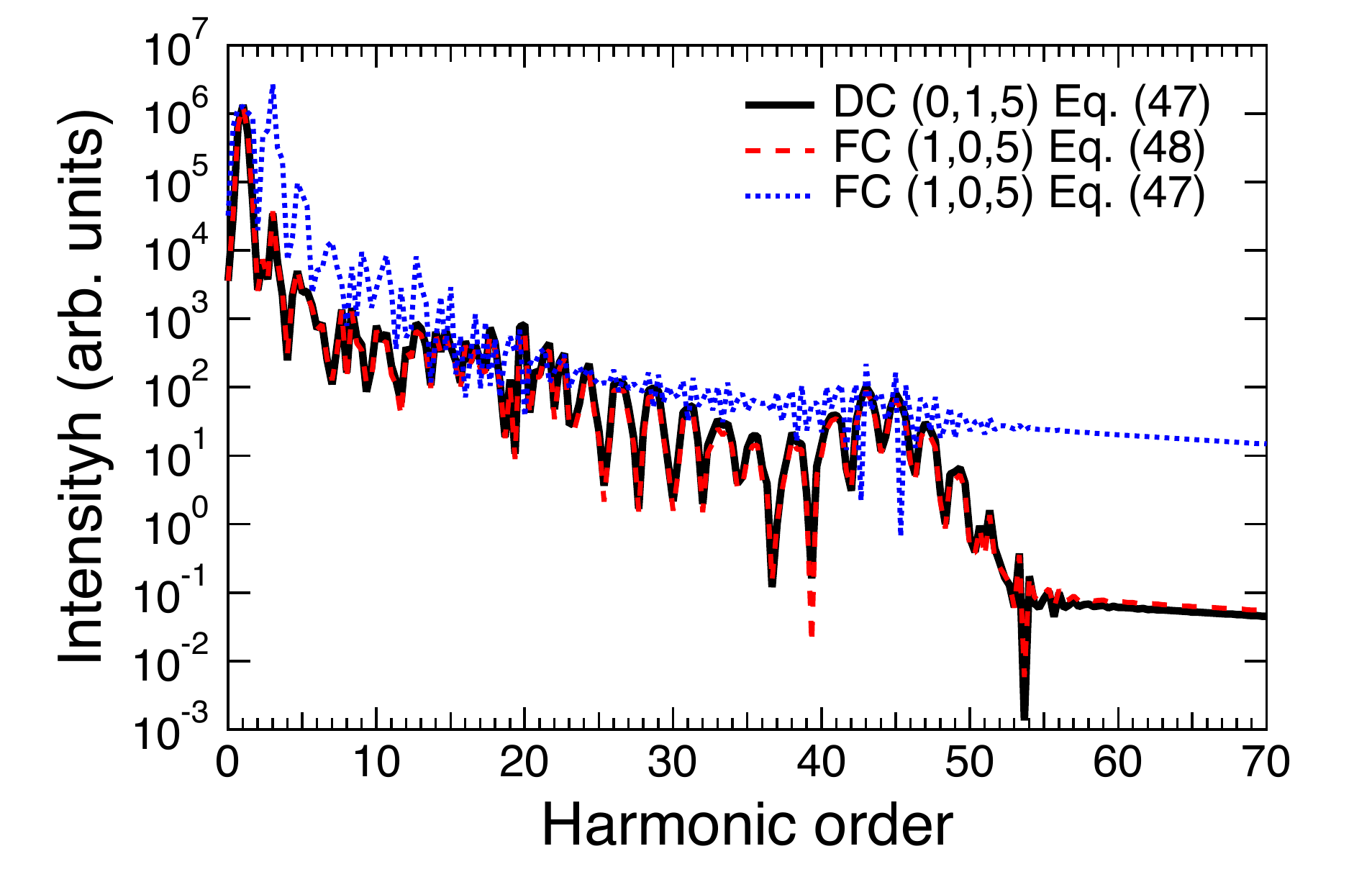}
	\end{center}
	\caption{HHG spectra of Be exposed to a laser pulse with a wavelength of 800 nm, an intensity of 3$\times$10$^{14}$ W/cm$^2$, and a foot-to-foot pulse width of three cycles. Comparison between the simulations with DC $(n_{fc}, n_{dc},n_{a})=(0,1,5)$ and FC $(1,0,5)$. For the case of FC, we also compare the results extracted via Eqs.~(\ref{eq:ehrenfest_acc}) and (\ref{eq:ehrenfest_fc_acc}).}
	\label{fig:BeHHG}
\end{figure}

Figure \ref{fig:BeHHG} compares the HHG spectra\index{High-harmonic generation (HHG)} from Be, calculated as the modulus squared of the Fourier transform of the dipole acceleration, extracted from the simulations with a dynamical and frozen core orbital.
If we calculate the frozen-core HHG spectrum using the modified formula Eq.~(\ref{eq:ehrenfest_fc_acc}), it overlaps with the DC result almost perfectly.
This indicates that the use of FC is a good approximation for the circumstances considered here.
However, the use of Eq.~(\ref{eq:ehrenfest_acc}) with FC leads to an erroneous spectrum.
Thus, it is essential to use Eq.~(\ref{eq:ehrenfest_fc_acc}) for the calculation of dipole acceleration and, then, HHG spectra from the simulation results with frozen core.

\begin{figure}[tb]
	\begin{center}
		\includegraphics[width=0.75\textwidth]{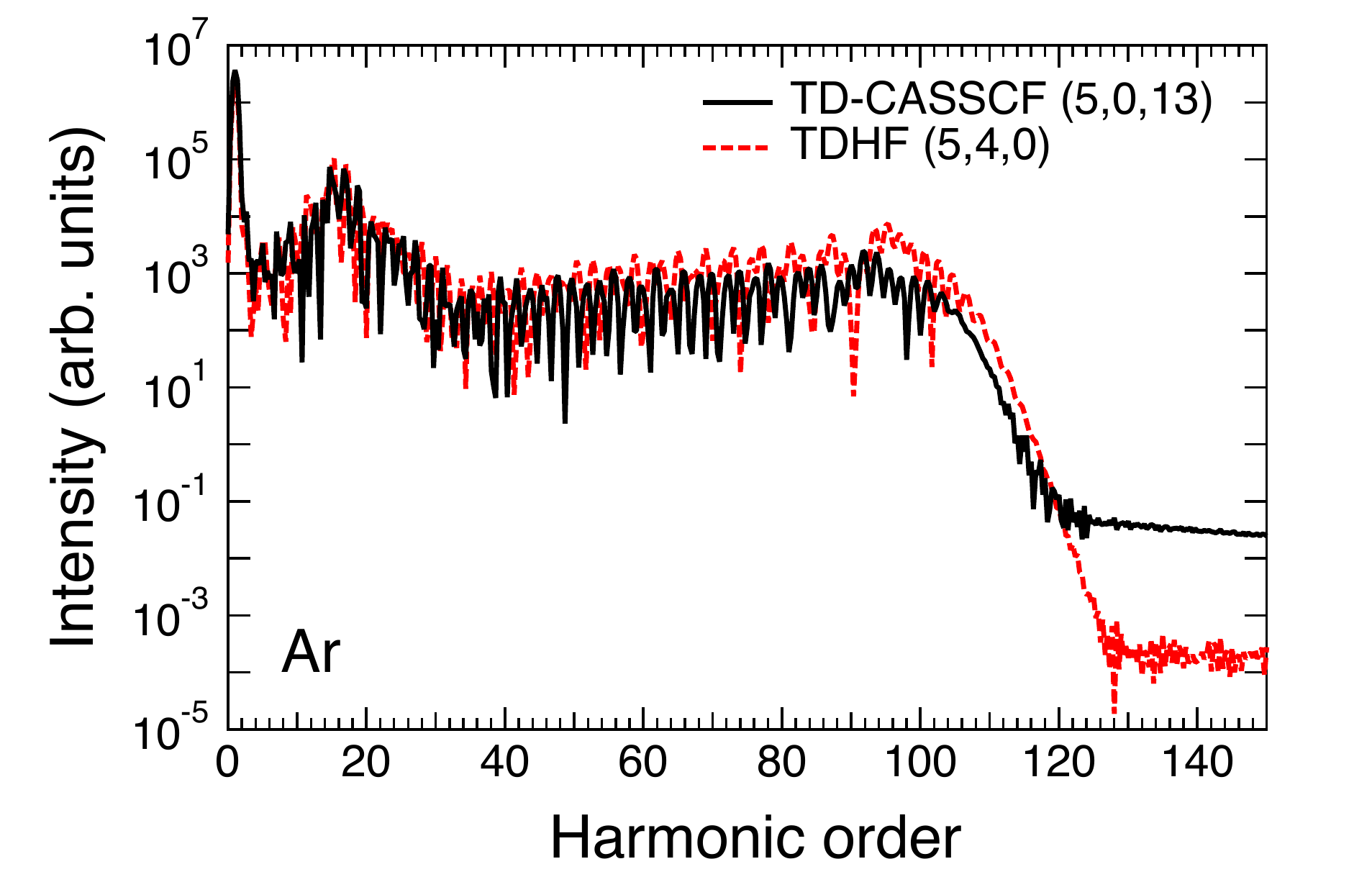}
	\end{center}
	\caption{HHG spectra of Ar exposed to a laser pulse with a wavelength of 800 nm, an intensity of 8$\times$10$^{14}$ W/cm$^2$, and a foot-to-foot pulse width of three cycles. Comparison between the TD-CASSCF $(n_{fc}, n_{dc},n_{a})=(5,0,13)$ and TDHF $(5,4,0)$.}
	\label{fig:ArHHG}
\end{figure}

\begin{figure}[tb]
	\begin{center}
		\includegraphics[width=0.75\textwidth]{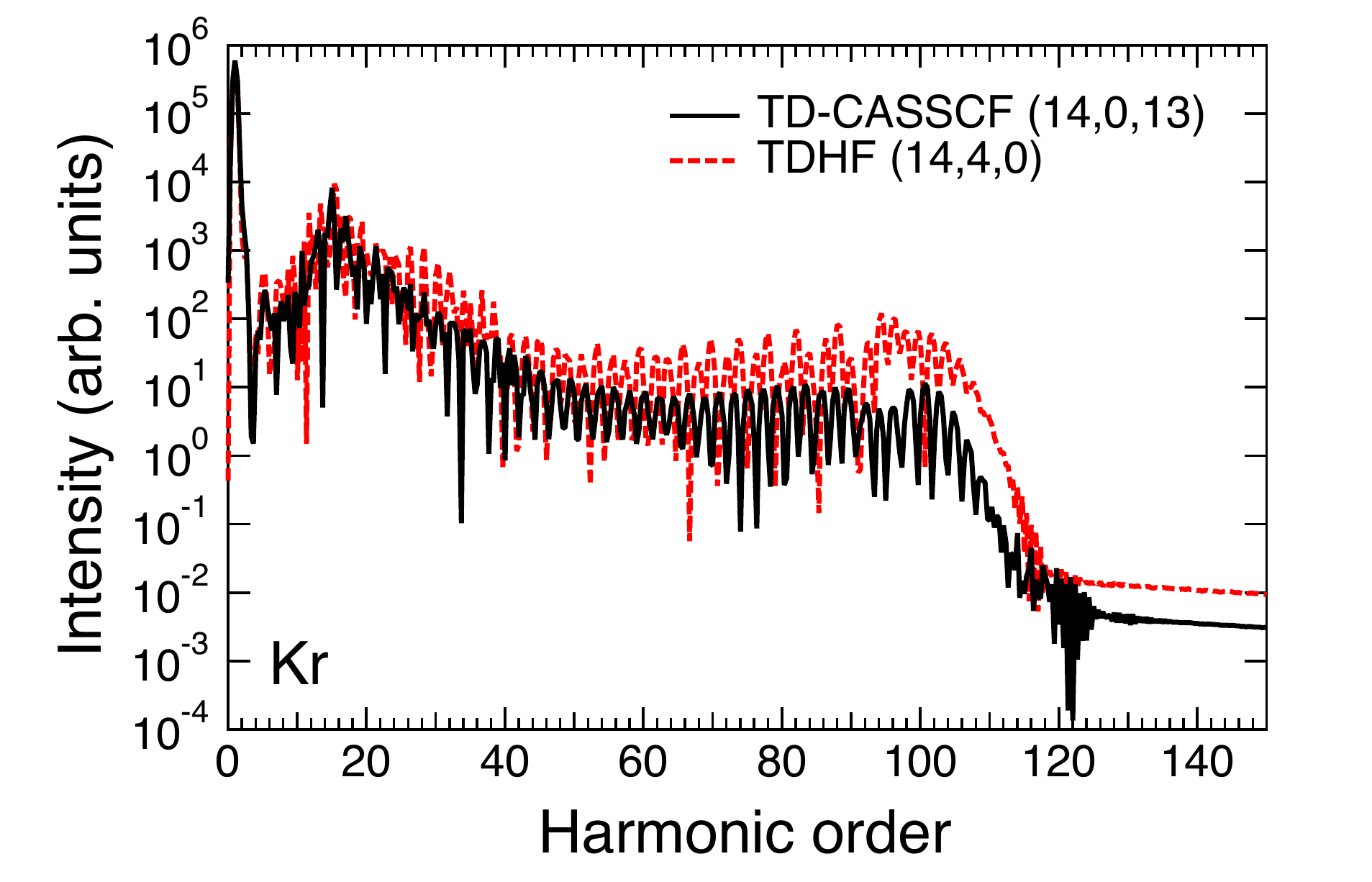}
	\end{center}
	\caption{HHG spectra of Kr exposed to a laser pulse with a wavelength of 800 nm, an intensity of 8$\times$10$^{14}$ W/cm$^2$, and a foot-to-foot pulse width of three cycles. Comparison between the TD-CASSCF $(n_{fc}, n_{dc},n_{a})=(14,0,13)$ and TDHF $(14,4,0)$.}
	\label{fig:KrHHG}
\end{figure}

We show in Fig.~\ref{fig:ArHHG} the HHG spectra of Ar calculated with the TD-CASSCF and TDHF methods.
These results well reproduce a dip around 53 eV ($\sim$34th order) that originates from the Cooper minimum and has been experimentally observed \cite{Worner2009PRL}. 
Whereas the difference between the TD-CASSCF and TDHF is not large in this case, it is more prominent in the HHG spectrum of Kr shown in Fig.~\ref{fig:KrHHG}; the TDHF overestimates the harmonic intensity near the cutoff more than one order of magnitude.
Such a quantitative difference is critical when we want to estimate the upper limit of the HHG pulse energy that can be generated with a given experimental setup.
It is wonderful that we can now achieve a converged simulation of high-harmonic generation from the thirty-six electron atom.

\subsection{Dipole Acceleration within the Single-Active-Electron Approximation}
\label{sec:DA within SAE}

The above discussion has important implications also for how to extract harmonic spectra from TDSE simulations of multielectron atoms and molecules within the single-active-electron approximation.
As stated above, Eq.~(\ref{eq:ehrenfest_acc}) with $\hat{V}_0$ replaced by the effective potential $V_{eff}$, corresponding to Eq.~(\ref{eq:force_fc}), is usually used:
\begin{eqnarray}\label{eq:ehrenfest_SAE_eff}
\langle a\rangle(t) = -
\langle\psi|\left(
\frac{\partial V_{eff}}{\partial z}+
\frac{\partial \hat{V}_{\rm ext}}{\partial z}
\right)|\psi\rangle.
\end{eqnarray}
On the other hand, Gordon {\it et al.} \cite{Gordon2006PRL} have argued that one should rather use Eq.~(\ref{eq:ehrenfest_acc}) as is, i.e., with the bare nuclear potential $\hat{V}_0$ ($=-\frac{Z}{r}$ for the atomic case):
\begin{eqnarray}\label{eq:ehrenfest_SAE_bare}
\langle a\rangle(t) = -
\langle\psi|\left(
\frac{\partial \hat{V}_0}{\partial z}+
\frac{\partial \hat{V}_{\rm ext}}{\partial z}
\right)|\psi\rangle.
\end{eqnarray}
They have taken the action-reaction law into account but ignored the binding force. 
However, the observation that Eq.~(\ref{eq:ehrenfest_fc_acc}) rather than Eq.~(\ref{eq:ehrenfest_acc}) has to be used in the presence of frozen-core orbitals, also numerically confirmed in Fig.~\ref{fig:BeHHG}, strongly suggests that, at the conceptual level, Eq.~(\ref{eq:ehrenfest_SAE_eff}) is the correct choice.

\subsection{Photoelectron energy spectrum}
\label{subsec:PES}

Time-resolved and angle-resolved photoelectron (photoemission) spectroscopy is becoming more and more important as a tool to probe ultrafast electron dynamics.
In principle, (angle-resolved) photoelectron energy spectrum can be calculated through projection of the departing wave packet onto plane waves or Coulomb waves (the difference in the results is usually negligibly small).
To apply this approach, however, we need to keep the wave function within the simulation box without being absorbed, which would lead to a huge computational cost.
As a new method that can be used with irECS, requiring a much smaller simulation box, the time-dependent surface flux (t-SURFF) method has recently been proposed \cite{Tao2012NJP}.
In this method, spectra are computed from the electron flux through a surface, beyond which the outgoing electron wave packet is absorbed by irECS. Instead of analyzing spectra at the end of the simulation, one can record the surface flux in the course of time evolution. We have recently succeeded in applying t-SURFF, originally formulated for SAE-TDSE simulations, to the TD-CASSCF simulations, whose details will be presented in a separate publication.

\begin{figure}[tb]
\begin{center}
\includegraphics[width=0.75\textwidth]{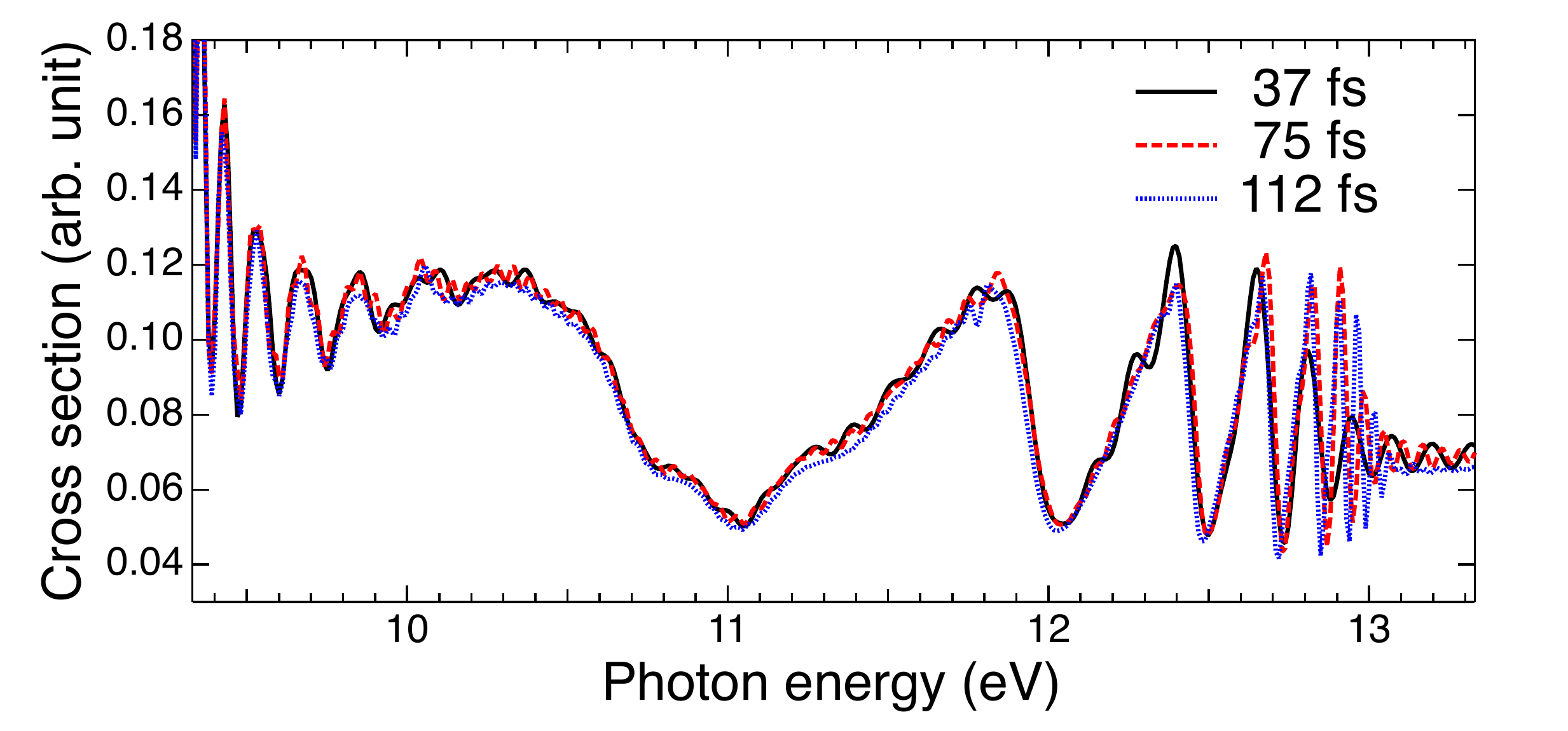} 	
\end{center} 
\caption{Relative photoionization cross section of Be as a function of photon energy extracted 37, 75, and 112 fs after the pulse from TD-CASSCF simulations with $(n_{fc},n_{dc},n_{a}) = (1,0,4)$ for a ultrabroadband three-cycle (foot-to-foot) pulse with 22 eV central photon energy.}
\label{fig:photoionization-cross-section-Be}
\end{figure}

Figure \ref{fig:photoionization-cross-section-Be} presents the calculated photoionization cross section of Be.
Making use of a broadband nature of an ultrashort pulse, one can draw such a plot with a single run, by dividing the photoelectron spectrum by photon energy spectrum.
The results, in reasonable agreement with reported measurements \cite{Wehlitz2003PRA}, well reproduce oscillating features due to the contribution from autoionizing states. 
We plot three curves extracted at different delays (37, 75, 112 fs) after the pulse.
We see that peaks grow around 13 eV with increasing delay, reflecting the evolution of autoionization.
Thus, the TD-CASSCF method can properly describe the process induced by electron correlation.

\begin{figure}[tb]
\begin{center}
\includegraphics[width=0.75\textwidth]{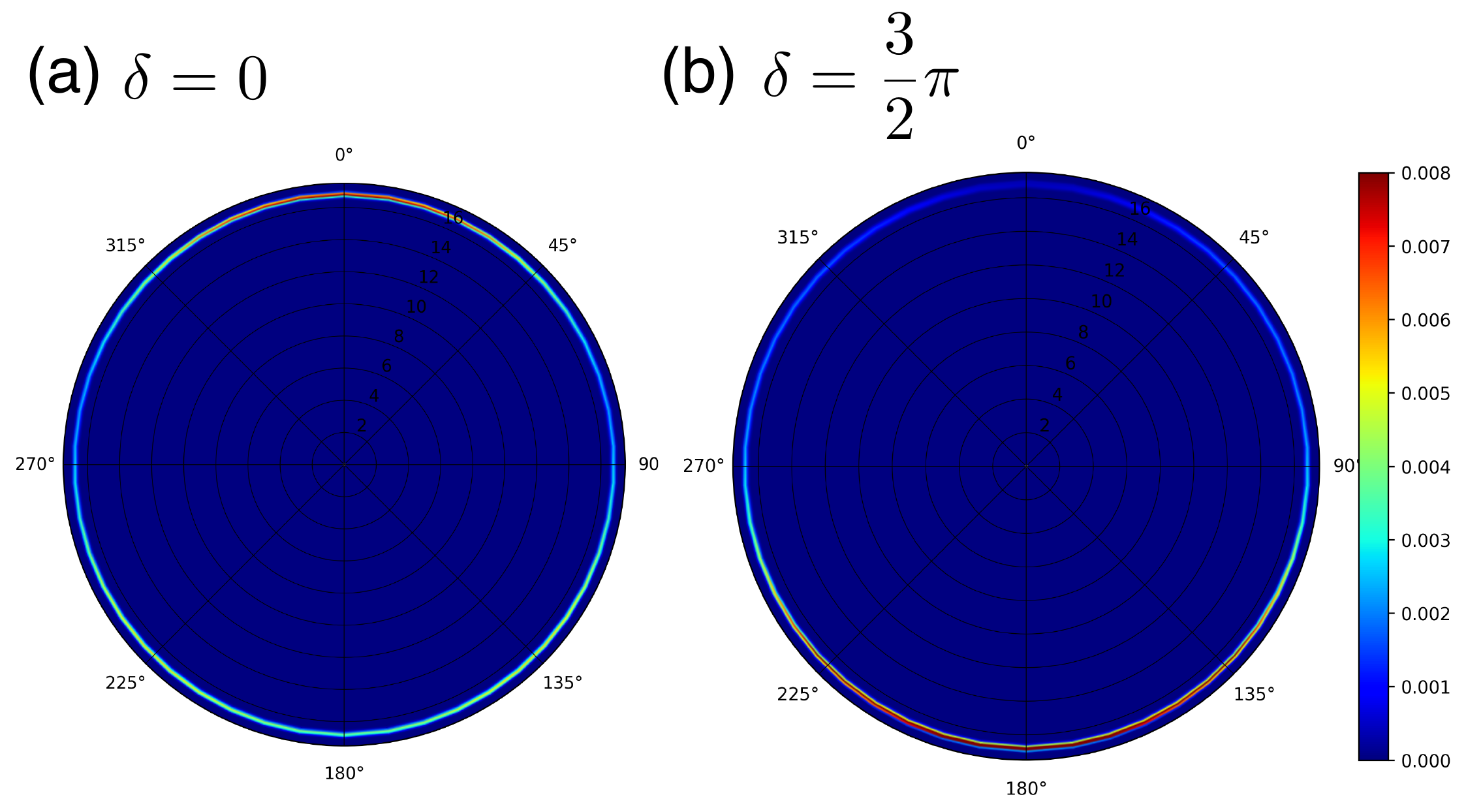} 	
\end{center} 
\caption{Angle-resolved photoelectron energy spectrum from Ne irradiated by a 10 fs bichromatic XUV pulse Eq.~(\ref{eq:bichromatic}) with $\omega=19.1\,{\rm eV}$ and (a) $\delta=0$ and (b) $\frac{3}{2}\pi$, calculated with the TDHF simulation. The $\omega$ and $2\omega$ intensities are $10^{13}\,{\rm W/cm}^2$ and $1.5\times 10^{9}\,{\rm W/cm}^2$, respectively.}
\label{fig:bichromatic-Ne}
\end{figure}

As a demonstration of capability to evaluate photoelectron angular distribution, let us consider a bichromatic XUV pulse whose electric field is of the form,
\begin{equation}
	E(t) = F_\omega (t) \cos\omega t + F_{2\omega} (t) \cos (2\omega t - \delta),
	\label{eq:bichromatic}
\end{equation}
where $F_\omega (t)$ and $F_{2\omega} (t)$ denote the envelopes of the $\omega$ and $2\omega$ pulses, respectively, and $\delta$ the relative phase.
Because of the interference between two-photon ionization by $\omega$ and single-photon ionization by $2\omega$, the photoelectron angular distribution is expected to vary with $\delta$.
This is confirmed by TDHF simulations as shown in Fig.~\ref{fig:bichromatic-Ne}.
Whereas roughly the same number of photoelectrons are emitted to the upper ($\sim 0^\circ$) and lower ($\sim 180^\circ$) hemispheres at $\delta=0$ (54\% to the lower hemisphere), approximately two-thirds (63 \%) of the electrons are emitted to the lower hemisphere at $\delta = \frac{3}{2}\pi$.
Hence, such simulations will be useful to design and analyze, e.g., coherent control experiments that can be realized by use of high-harmonic and free-electron-laser XUV sources with temporal coherence \cite{Prince2016NPhoton,Iablonskyi2017PRL}.

\section{Summary}
\label{sec:Summary}

We have compiled our recent development of the time-dependent complete-active-space self-consistent-field method to simulate multielectron dynamics in ultrafast intense laser fields along with numerical examples for atoms.
Introducing the concept of frozen core, dynamical core, and active orbital subspace decomposition, TD-CASSCF allows compact and, at the same time, accurate representation of correlated multielectron dynamics in strongly driven atoms and molecules.
It also has desirable features of gauge invariance and size extensivity.
We can now handle strong-field phenomena in systems containing tens of electrons from the first principles, which was merely a dream several years ago.

While the present work has focused on the TD-CASSCF method, especially, for atoms, we have developed and been actively developing a variety of different {\it ab initio} methods.
We have numerically implemented the MCTDHF method for molecules, based on a multiresolution Cartesian grid, without need to assume any symmetry of molecular structure \cite{Sawada2016PRA}.
We have developed the TD-ORMAS method \cite{Sato2015PRA}, which is more approximate and thus computationally even less demanding than TD-CASSCF, and allows one to handle general MCSCF wave functions with arbitrary CI spaces.
We have more recently formulated the time-dependent optimized coupled-cluster method \cite{Sato2018JCP}, based not on multiconfiguration expansion but on coupled-cluster expansion.
This method is gauge invariant, size extensive, and polynomial cost-scaling.
Furthermore, as an alternative that can in principle take account of correlation effects and extract any one- and two-particle observable while bypassing explicit use of the wave function, we have reported a numerical implementation of the time-dependent two-particle reduced density matrix method \cite{Lackner2015PRA,Lackner2017PRA}.
Whereas the above methods concentrate on the electron dynamics, we have also considered electron-nuclear dynamics and formulated a fully general TD-MCSCF method to describe the dynamics of a system consisting of arbitrary different kinds and numbers of interacting fermions and bosons \cite{Anzaki2017PCCP}.
All these developments will open various, flexible new possibilities of highly accurate {\it ab initio} investigations of correlated multielectron and multinucleus quantum dynamics in ever-unreachable large systems.

\begin{acknowledgement}
This research was supported in part by a Grant-in-Aid for Scientific Research (Grants No. 23750007, No. 23656043, No. 23104708, No. 25286064, No. 26390076, No. 26600111, No. 16H03881, and 17K05070) from the Ministry of Education, Culture, Sports, Science and Technology (MEXT) of Japan and also by the Photon Frontier Network Program of MEXT. 
This research was also partially supported by the Center of Innovation Program from the Japan Science and Technology Agency, JST, and by CREST (Grant No. JPMJCR15N1), JST. 
Y.~O.~gratefully acknowledges support from the Graduate School of Engineering, The University of Tokyo, Doctoral Student Special Incentives Program (SEUT Fellowship). 
O.~T.~gratefully acknowledges support from the Japanese Government (MEXT) Scholarship. 
We thank I. B\v{r}ezinov\'a, F. Lackner, S. Nagele, J. Burgd\"orfer, and A. Scrinzi for fruitful collaborations that have greatly contributed to this work.
\end{acknowledgement}

\printindex

\end{document}